# Transition operator approach for the description of spontaneous decay in a multi-qubit system

**Ya. S. Greenberg and O. A. Chuikin**

Novosibirsk State Technical University, 630073 K. Marx Ave. 20,
Novosibirsk, Russia

**Abstract**

In this paper we discuss the use of the transition operator method for the theoretical description of a multi-qubit system in a one-dimensional waveguide. A general calculation has been performed for the N-qubit system, which was then applied to the case of spontaneous decay for one and two qubits. The probabilities of transitions in such systems, as well as the emission spectra, are investigated in detail.

## 1. Introduction

Circuit quantum electrodynamics (QED) is a very promising field of study that allows us to closely look at the physics of light-matter interaction on a quantum scale [1]. It has a great number of applications such as quantum information processing [2-5], single-photon transistor [6-8], quantum metamaterials [9, 10], microwave photonics [11], and many other important research fields. Superconducting qubits coupled to microwave photons in one-dimensional (1D) waveguides [12-14] are in the core of circuit QED. Due to spatial confinements, this setup allows one to achieve an almost ideal mode matching [15] which results in a strong coupling and even ultrastrong coupling regimes when the interaction strength overwhelms relaxation rates [16, 17]. These kinds of experiments are very challenging to perform for regular atoms in optical domain.

Another significant difference between natural atoms and superconducting qubits is that in circuit QED we can create artificial atoms with desirable parameters some of which can be tuned, e.g. resonant frequency or coupling strength [16], and, what is more important, we can address and manipulate the artificial atoms individually [18]. This makes it possible to in-depth study of different types of interaction between a few qubits which can be far more interesting and complex than just one qubit in a cavity [19, 20]. Because of the perfect mode matching the exchange interaction between the qubits is very strong and critically depends on the effective distance between them which can be tuned by changing the wavelength via qubit resonant frequency. Different spatial placements of qubits in a chain can lead to significant modification of decay rates [21, 22]. This collective effect can induce radiative decay and result in superradiance [23-25], or decrease the decay rates, which corresponds to subradiance [26-28]. Other important research works on many-qubit systems include the creation of entanglement both with microwave photons [29, 30] and the qubits themselves [31, 32] which is particularly useful for the creation of quantum gates [5]. Despite a large number of works on multi-qubit systems, the theoretical description of dynamics of individual qubits in a multi-qubit system are still quite non-trivial and has many open problems.

Here we present a new method for the description of transitions and spontaneous decay in multi-qubit systems interacting with microwave photons in a 1D transmission line. Our approach is based on the so-called transition operator, firstly introduced by Lehmberg [33] to theoretically describe the spontaneous emission of two-level atoms in free space [34, 35].

The exact expressions for the matrix elements of the transition operators can be obtained within standard quantum mechanics formalism using Heisenberg equations. The tracing out the photonic modes from equations of motion allows us to obtain equations only for atomic operators independent of the photon number. As distinct from conventional density matrix approach, differential equations for the matrix elements of the transition operator are linear and therefore easier to solve analytically. Moreover, because the solutions of equations are operator functions, they are independent of the specific initial state of the system. It means that we don't need to find a new solution for every new initial state like in the case of equations for the density matrix. Once the exact solutions for transition operators are found, we can use them not only to obtain transition probabilities but also for the calculation of the photonic emission spectrum which requires only the knowledge of the initial density matrix. As we show here, transition operators and elements of the density matrix are very closely related, so one can easily switch from one approach to the other one if needed.

We also should note the other methods of description of 1D qubit chains. One can use input-output formalism [21] which requires the density matrix equations. The other approach includes direct calculations of probability amplitudes for single-excitation wavefunction, as was done in [36]. Even though these methods provide similar results for some calculations, the transition operator approach is a more general way that allows for obtaining general equations for N-qubits beyond a single-photon approximation.

The paper is organized as follows. In Sec. 2 we define the transition operator, describe its general properties, and establish its connection to a density matrix. The expression of transition probabilities between qubit states in terms of the transition operator is also shown. In Sec. 3 we show the application of the method for the description of N-qubit system in a 1D transmission line. Using a Jaynes-Cummings Hamiltonian we obtain Heisenberg differential equations for the matrix elements of the transition operator by tracing out the photon field. Finally, performing Markov approximation we obtain linear differential equations of motion for arbitrary transition operators from which the matrix elements of the transition operator can be found. In Sec. 4 we present a general expresssion for calculating the spectral density of photons using two-time correlation functions of spin operators. We also show how these types of correlation functions can be calculated in terms of the transition operators and initial density matrix. This approach allows us to circumvent the quantum regression theorem [37, 38], the approximation usually used to find the two-time correlation functions. In Sec. 5 we apply our method to the simplest case of one qubit in a coplanar waveguide and find transition probabilities and photon spectrum in terms of the matrix elements of the transition operator. In Sec. 6 we consider a more complex system which contains two qubits in a waveguide. Using the transition operator approach we derive transition probabilities for spontaneous decay from different initial states including two excited qubits, one excited qubit, and symmetric and asymmetric entangled states. Moreover, we calculate the photon spectrum for the two-qubit system in general form and then specify the solution for different initial states which include symmetric and asymmetric entangled states, states with one excited qubit with the other qubit being in a ground state, and the state with both qubits being excited. Additionally, we consider the initial states when one or two of the qubits can be prepared in a superposition state, and show that spectra for these states can be expressed as a combination of previously obtained spectra. A summary of our work and some clothing thoughts are presented in Sec. 7.

## 2. General properties of the transition operator

We consider a system of N identical qubits with eigenstates $|i\rangle$ coupled to a continuum of photon modes $|v\rangle$ in a one-dimensional open waveguide. Our main interest is the probability of transition from some arbitrary initial state of a qubit system $|\Psi_0\rangle$ with no photons to some final state $|\Psi_1\rangle$ with v photons in the field. According to the general principles of quantum mechanics, the probability amplitude of such transition is given by the following matrix element:

$$\langle \Psi_1, v | e^{-iHt} | \Psi_0, 0 \rangle, \qquad (2.1)$$

where H is the complete Hamiltonian which includes the N-qubit system, photon field, and their interaction, $|0\rangle, |v\rangle$ are the Fock states with zero and v photons, respectively. To find the total transition probability we must find the squared modulus of this amplitude and sum it over the complete set of possible final photon states $|\mu\rangle$ for the field. The expression for the completeness of the set of the photon states $|\mu\rangle$ is $\sum_\mu |\mu\rangle\langle\mu| = 1$. Therefore we then obtain a total transition probability:

$$W_{0\to 1} = \sum_\mu \left|\langle \Psi_1, \mu | e^{-iHt} | \Psi_0, 0 \rangle\right|^2 = \sum_\mu \langle \Psi_0, 0 | e^{iHt} | \Psi_1, \mu \rangle\langle \Psi_1, \mu | e^{-iHt} | \Psi_0, 0 \rangle = \langle \Psi_0 | \langle 0 | e^{iHt} | \Psi_1 \rangle\langle \Psi_1 | e^{-iHt} | 0 \rangle | \Psi_0 \rangle \qquad (2.2)$$

The wave function $|\Psi_1\rangle$ can be expanded over the complete set of qubit states:

$$|\Psi_1\rangle = \sum_i c_i |i\rangle. \qquad (2.3)$$

Then we can rewrite (2.2) in the following form:

$$W_{0\to 1} = \sum_{i,j} c_i c_j^* \langle \Psi_0 | \langle e^{iHt} |i\rangle\langle j| e^{-iHt} \rangle_0 | \Psi_0 \rangle. \tag{2.4}$$

where $\langle ...... \rangle_0$ is the average over photon vacuum $\langle ...... \rangle_0 = \langle 0| ....... |0\rangle$.

Following Lehmberg [33], we define a transition operator:

$$P_{i,j}(t) = e^{iHt} |i\rangle\langle j| e^{-iHt}. \tag{2.5}$$

The expression (2.4) can then be written in a form:

$$W_{0\to 1} = \sum_{i,j} c_i c_j^* \langle \Psi_0 | \langle P_{i,j}(t) \rangle_0 | \Psi_0 \rangle. \tag{2.6}$$

Thus, the probability of transition from the eigenstate $|n\rangle$ to the eigenstate $|m\rangle$ can be calculated by using the corresponding matrix element of the transition operator $P_{m,m}$ averaged over photon vacuum:

$$W_{nm} = \langle n | \langle P_{m,m} \rangle_0 | n \rangle. \tag{2.7}$$

It follows from the completeness of the qubits states $|i\rangle$ that the sum of diagonal elements of the transition operator is equal to one:

$$\sum_i P_{i,i} = 1 \tag{2.8}$$

As follows from the definition (2.5), the expressions for the elements of the transition operator satisfy the Heisenberg equation:

$$\frac{d}{dt} P_{i,j}(t) = i [H, P_{i,j}(t)]. \tag{2.9}$$

with the initial conditions $P_{i,j}(0) = |i\rangle\langle j|$.

Unlike the equations for spin operators or elements of the density matrix, equations for $P_{i,j}$ are linear: they contain only the first degrees of the same operators. The number of equations for the transition operator is determined by the number of states from the complete set. For example, for one qubit there are only two states and, accordingly, four equations: for $P_{ee}$, $P_{gg}$, $P_{eg}$, and $P_{ge}$. However, there are only three independent equations since the $P_{eg}$ is a complex conjugate to a $P_{ge}$. For two qubits there are four states and, consequently, there are ten independent equations – four equations for the diagonal elements of the transition operator and six equations for off-diagonal ones (excluding complex conjugates).

In the general case, the solution of the equations of motion for elements of the transition operator averaged over photon vacuum has the following form:

$$\langle P_{i,j}(t) \rangle_0 = \sum_{m,n} c_{mn}^{ij}(t) |m\rangle\langle n|, \tag{2.10}$$

where $c_{mn}^{ij}(t)$ are c- numbers.

By the definition, the qubit density matrix of a spin system, $\rho_{S,ji}(t) = \langle j | Tr_V [\rho(t)] | i \rangle \equiv \langle j | \rho_S(t) | i \rangle$, where $\rho(t)$ is the density matrix of the whole system:

$$\rho(t) = e^{-iHt} \rho(0) e^{iHt}, \tag{2.11}$$

can be expressed in terms of the average value of the transition operator:

$$\rho_{S,ji}(t) = Tr_{S,v} (\rho(0) P_{i,j}(t)) \equiv \langle P_{i,j}(t) \rangle, \tag{2.12}$$

where the trace is taken over both the qubit system and the photon field and $\rho(0)$ is the initial density matrix of the whole system. If we assume that initially the field is in a photon vacuum:

$$\rho(0) = \rho_S(0) \times \rho_V(0) = \rho_S(0) \times |0\rangle\langle 0|, \tag{2.12a}$$

we then obtain from (2.12):

$$\langle l | \rho_S(t) | m \rangle = \sum_{n.,q} \langle n | \rho_S(0) | q \rangle \langle q | \langle 0 | P_{m,l}(t) | 0 \rangle | n \rangle. \tag{2.13}$$

Thus, knowing the expressions for the vacuum average of the matrix elements of the operator $P_{i,j}(t)$ we can easily find the corresponding values of the matrix elements $\rho_{S,ij}(t)$ if the initial density matrix $\rho^{(s)}(0)$ is known. It should be noted that in contrast to the elements of the density matrix $\rho_{S,ij}(t)$, which are numerical functions, the matrix elements $P_{i,j}(t)$ are the operator functions.

## 3. General case of N qubits

Consider a system consisting of N qubits in a one-dimensional infinite waveguide. This system can be described by a Jaynes-Cummings Hamiltonian:

$$H = \frac{1}{2}\sum_{n=1}^{N}\left(1+\sigma_z^{(n)}\right)\Omega_n + \sum_k \omega_k a_k^\dagger a_k + \sum_k \left(a_k^\dagger S_k^- + S_k^+ a_k\right), \quad (3.1)$$

where we introduced collective atomic spin operators:

$$S_k^- = \sum_{n=1}^{N} g_k^{(n)} e^{-ikx_n} \sigma_-^{(n)}, \qquad S_k^+ = \sum_{n=1}^{N} g_k^{*(n)} e^{ikx_n} \sigma_+^{(n)}. \quad (3.2)$$

Here $\sigma^{(n)}{}_z$ is a Pauli spin operator and $\Omega_n$ is a resonant frequency of n-th qubit, $a_k^\dagger$ ($a_k$) are a creation (annihilation) operators for a photon with $k$ mode, $\omega_k$ is a photon frequency, $\sigma_-^{(n)} = |g\rangle_n{}_n\langle e|$ and $\sigma_+^{(n)} = |e\rangle_n{}_n\langle g|$ are the atomic "ladder" operators which lower or raise a state of the n-th qubit, $g_k$ is a coupling strength between the qubit and the field, $x_n$ is a spatial coordinate of the n-th qubit.

From (2.9) and (3.1) we obtain the equation of motion for the transition operator:

$$\begin{aligned}\frac{dP_{ij}}{dt} &= i\frac{1}{2}\sum_{n=1}^{N}\Omega_n\left(e^{iHt}\sigma_Z^{(n)}|i\rangle\langle j|e^{-iHt} - e^{iHt}|i\rangle\langle j|\sigma_Z^{(n)}e^{-iHt}\right) \\ &+ i\sum_k a_k^\dagger(t)\left(e^{iHt}S_k^-|i\rangle\langle j|e^{-iHt} - e^{iHt}|i\rangle\langle j|S_k^-e^{-iHt}\right) \\ &+ i\sum_k \left(e^{iHt}S_k^+|i\rangle\langle j|e^{-iHt} - e^{iHt}|i\rangle\langle j|S_k^+ e^{-iHt}\right)a_k(t)\end{aligned} \quad (3.3)$$

where the photonic operators are in the Heisenberg representation:

$$a_k^\dagger(t) = e^{iHt}a_k^\dagger e^{-iHt}, \quad a_k(t) = e^{iHt}a_k e^{-iHt}. \quad (3.4)$$

For photonic operators we obtain the equations of motion from the Heisenberg equation:

$$i\frac{da_k}{dt} = [a_k(t), H] = \omega_k a_k(t) + S_k^-(t), \quad (3.5a)$$

$$i\frac{da_k^\dagger}{dt} = \left[a_k^\dagger(t), H\right] = -\omega_k a_k^\dagger(t) - S_k^+(t). \quad (3.5b)$$

where $S_k^\pm(t)$ are collective spin operators in the Heisenberg picture. The formal solution of these equations is given by

$$a_k(t) = a_k(0)e^{-i\omega_k t} - i\int_0^t e^{-i\omega_k(t-\tau)}S_k^-(\tau)d\tau, \quad (3.6a)$$

$$a_k^\dagger(t) = a_k^\dagger(0)e^{i\omega_k t} + i\int_0^t e^{i\omega_k(t-\tau)}S_k^+(\tau)d\tau. \quad (3.6b)$$

Even though the photon operators $a_k^\dagger(t)$, $a_k(t)$ commute with collective spin operators $S_k^\pm(t)$, each term in (3.6a) and (3.6b) does not commute with $S_k^\pm(t)$. This explains the position of these operators in the third term of Hamiltonian (3.1). They should be placed in such a way that in the final expression the creation operators $a_k^\dagger(0)$ were placed on the left side of the transition operator $P_{i,j}$, while the annihilation operators $a_k(0)$ were placed on the right side of the transition operator. This is necessary for the corresponding terms to be dropped out upon averaging over the photon vacuum in (2.10).

Substituting the expressions (3.6) into the equation of motion (3.3) we obtain:

$$\begin{aligned}\frac{dP_{ij}}{dt} = & i\frac{1}{2}\sum_{n=1}^{N}\Omega_n\left(e^{iHt}\sigma_Z^{(n)}|i\rangle\langle j|e^{-iHt} - e^{iHt}|i\rangle\langle j|\sigma_Z^{(n)}e^{-iHt}\right)\\ & +i\sum_k a_k^\dagger(0)e^{i\omega_k t}e^{iHt}S_k^-|i\rangle\langle j|e^{-iHt} - i\sum_k a_k^\dagger(0)e^{i\omega_k t}e^{iHt}|i\rangle\langle j|S_k^- e^{-iHt}\\ & +i\sum_k e^{iHt}S_k^+|i\rangle\langle j|e^{-iHt}a_k(0)e^{-i\omega_k t} - i\sum_k e^{iHt}|i\rangle\langle j|S_k^+ e^{-iHt}a_k(0)e^{-i\omega_k t}\\ & +\sum_k \int_0^t e^{i\omega_k(t-\tau)}S_k^+(\tau)d\tau\, e^{iHt}\left[|i\rangle\langle j|,S_k^-\right]e^{-iHt} + \sum_k e^{iHt}\left[S_k^+,|i\rangle\langle j|\right]e^{-iHt}\int_0^t e^{-i\omega_k(t-\tau)}S_k^-(\tau)d\tau\end{aligned} \quad ; \quad (3.7)$$

Here we write the equation in such a manner that the action of atomic operators on system states is clearly visible. Equation (3.7a) can be rewritten in terms of transition operators and atomic operators in the Heisenberg picture, since $e^{iHt}\sigma_Z^{(n)}|i\rangle\langle j|e^{-iHt} = e^{iHt}\sigma_Z^{(n)}e^{-iHt}e^{iHt}|i\rangle\langle j|e^{-iHt} = \sigma_Z^{(n)}(t)P_{i,j}(t)$ (the same procedure applies to the terms with $S_k^\pm$ as well).

Up to now, we did not make any approximations: the above expressions are exact. In order to solve the equation (3.7), the following assumptions are made:

$$\sigma_-^{(n)}(\tau) \approx \sigma_-^{(n)}(t)e^{-i\Omega_n(\tau-t)}, \tag{3.8a}$$

$$\sigma_+^{(n)}(\tau) \approx \sigma_+^{(n)}(t)e^{i\Omega_n(\tau-t)}, \tag{3.8b}$$

Assuming that all resonant frequencies of qubits are identical and equal to some value $\Omega$, we then obtain:

$$S_k^-(\tau) \approx S_k^-(t)e^{-i\Omega(\tau-t)}, \tag{3.9a}$$

$$S_k^+(\tau) \approx S_k^+(t)e^{i\Omega(\tau-t)}. \tag{3.9b}$$

As is proven in [21, 34] the assumptions (3.8), (3.9) are equivalent to Wigner-Weisskopf or Markov approximations. It allows us to take $S_k^\pm(\tau)$ out of the integrand in (3.7b). We then rewrite obtained integral functions by taking into account the resonant approximation, i.e. assuming that the integral makes the main contribution near the resonance frequency $\Omega$. This allows us to take the upper limit to infinity, and we get:

$$\int_0^t e^{i(\omega-\Omega)(t-t')}dt' \approx \int_0^\infty e^{i(\omega-\Omega)\tau}d\tau = \pi\delta(\omega-\Omega) + i\,P.v.\left(\frac{1}{\omega-\Omega}\right), \tag{3.10}$$

where $\delta(\omega)$ is a Dirac delta function and P.v. is a Cauchy principal value.

According to all assumptions above, we obtain from (3.7) the equation of motion for the transition operator in the following form:

$$\begin{aligned}\frac{dP_{ij}}{dt} = & i\frac{1}{2}\sum_{n=1}^{N}\Omega_n\left(e^{iHt}\sigma_Z^{(n)}|i\rangle\langle j|e^{-iHt} - e^{iHt}|i\rangle\langle j|\sigma_Z^{(n)}e^{-iHt}\right)\\ & +i\sum_k a_k^\dagger(0)e^{i\omega_k t}e^{iHt}\left[S_k^-,|i\rangle\langle j|\right]e^{-iHt} + i\sum_k e^{iHt}\left[S_k^+,|i\rangle\langle j|\right]e^{-iHt}a_k(0)e^{-i\omega_k t}\\ & +\sum_{n,m}^{N}\frac{\Gamma_{n,m}}{2}e^{iHt}\left(2\sigma_+^{(m)}|i\rangle\langle j|\sigma_-^{(n)} - \sigma_+^{(m)}\sigma_-^{(n)}|i\rangle\langle j| - |i\rangle\langle j|\sigma_+^{(m)}\sigma_-^{(n)}\right)e^{-iHt}\\ & +i\sum_{n,m}^{N}\alpha_{n,m}e^{iHt}\left(|i\rangle\langle j|\sigma_+^{(m)}\sigma_-^{(n)} - \sigma_+^{(m)}\sigma_-^{(n)}|i\rangle\langle j|\right)e^{-iHt}\end{aligned} \quad ; \quad (3.11)$$

where according to the Fermi Golden rule we have introduced a decay rate $\Gamma_{n,m}$

$$\Gamma_{n,m} = \sum_k 2\pi\delta(\omega_k - \Omega)g_k^{(n)}g_k^{*(m)}e^{-ik(x_n-x_m)}, \tag{3.12a}$$

and a frequency shift $\alpha_{n,m}$:

$$\alpha_{n,m} = \sum_k g_k^{(n)}g_k^{*(m)}e^{-ik(x_n-x_m)}P.v.\left(\frac{1}{\omega_k - \Omega}\right). \tag{3.12b}$$

Equation (3.11) is the most general case of the equation of motion for the matrix elements of the transition operator (2.5) for N qubits with identical resonant frequencies. When equation (3.11) is averaged over the initial photon vacuum, the second line in (3.11) can be dropped out.

For a long 1D waveguide we can replace the summation over k by the integration:

$$\sum_k \to \frac{L}{2\pi}\int_{-\infty}^{+\infty} d|k| = \frac{L}{\pi v_g}\int_0^\infty d\omega, \qquad (3.13)$$

where L is the quantization length in the propagation direction and $v_g$ is the group velocity.

If we assume that the coupling strength is the same for all qubits, $g_k^{(n)} = g_k^{(m)} \equiv g_k$ and is symmetrical, i.e. $g_k = g_{-k}$, and it contributes mainly near the resonance $g_k \approx g_{k_0}$, where $k_0 = \Omega/v_g$, we then obtain for (3.12a) and (3.12b):

$$\Gamma_{n,m} = \Gamma \cos(k_0 |d_{n,m}|), \qquad (3.14a)$$

$$\alpha_{n,m}(k) = -\frac{\Gamma}{2}\sin(k_0 |d_{n,m}|). \qquad (3.14b)$$

where

$$\Gamma = \frac{2L}{v_g}|g_{k_0}|^2 \qquad (3.14c)$$

and $d_{n,m} = (x_n - x_m)$ is the distance between the n-th and an m-th qubit. The expression (3.14b) is obtained with the help of the following relation [39]:

$$P.v.\int_0^{+\infty} \frac{\cos(\omega d_{n,m}/v_g)}{\omega - \Omega} d\omega = -\pi \sin(k_0 |d_{n,m}|). \qquad (3.15)$$

## 4. Radiation spectrum and calculation of the correlation functions

In circuit quantum electrodynamics it is possible to experimentally measure both the full photon spectrum $\langle a_k^\dagger(t) a_k(t) \rangle$ and one-time mean values of single-photon operators $\langle a_k(t) \rangle, \langle a_k^+(t) \rangle$ [30]. Moreover, we can construct more complex photon correlation functions with a higher order of photon operators and experimentally measure them as well [30].

From the equations for the photon operators (3.6) we can calculate the spectral density of electromagnetic radiation by averaging either $\langle a_k^\dagger(t) a_k(t) \rangle$ or $\langle a_k(t) \rangle, \langle a_k^+(t) \rangle$ over photon vacuum (see equation (2.12)):

$$\langle a_k^\dagger(t) \rangle = i\int_0^t d\tau\, e^{i\omega_k(t-\tau)} \langle S_k^+(\tau) \rangle, \qquad (4.1a)$$

$$\langle a_k(t) \rangle = -i\int_0^t d\tau\, e^{-i\omega_k(t-\tau)} \langle S_k^-(\tau) \rangle. \qquad (4.1b)$$

$$\langle a_k^\dagger(t) a_k(t) \rangle = \int_0^t d\tau \int_0^t d\tau'\, e^{-i\omega_k(\tau-\tau')} \langle S_k^+(\tau) S_k^-(\tau') \rangle, \qquad (4.2)$$

As can be seen from these equations, the spectra are proportional either to two-time atomic correlation functions $\langle S_k^+(\tau) S_k^-(\tau') \rangle$ or to the one-time mean values, $\langle S_k^\pm(\tau) \rangle$. To calculate these quantities we can use the transition operators obtained from the equation (3.11). This method is described below.

Let us start with a function $\langle S_k^+(t) \rangle$, which reads (see equation (3.2).

$$\langle S_k^+(t) \rangle = \sum_n^N g_k^{*(n)} e^{ikx_n} \langle \sigma_+^{(n)}(t) \rangle. \qquad (4.3)$$

where:

$$\langle \sigma_+^{(n)}(t) \rangle = Tr_{S,\nu}\left[\sigma_+^{(n)}(t)\rho(0)\right], \qquad (4.4)$$

In (4.3) the trace is taken over both the atomic states S and photon states ν, and ρ(0) is the initial density matrix of the whole system. If the initial photon state is a vacuum as is given in (2.12a), we can rewrite (4.4) as follows:

$$\langle \sigma_+^{(n)}(t) \rangle = Tr_S\left[\langle 0|\sigma_+^{(n)}(t)|0\rangle \rho_S(0)\right] = Tr_S\left[\langle \sigma_+^{(n)}(t) \rangle_0 \rho_S(0)\right]. \qquad (4.5)$$

Therefore, for (4.1a), (4.1b) we obtain:

$$\left\langle a_k^\dagger(t) \right\rangle = i\sum_n^N g_k^{*(n)} e^{ikx_n} \int_0^t d\tau e^{i\omega_k(t-\tau)} Tr_s\left(\left\langle \sigma_+^{(n)}(\tau)\right\rangle_0 \rho_S(0)\right), \tag{4.6a}$$

$$\left\langle a_k(t) \right\rangle = i\sum_n^N g_k^{(n)} e^{-ikx_n} \int_0^t d\tau e^{-i\omega_k(t-\tau)} Tr_s\left(\left\langle \sigma_-^{(n)}(\tau)\right\rangle_0 \rho_S(0)\right). \tag{4.6b}$$

For N-qubit system spin operators $\sigma_+^{(n)}(t)$, $\sigma_-^{(n)}(t)$ for an $n$th spin can be expressed in the form of a transition operator for an $n$th spin:

$$\left\langle \sigma_+^{(n)}(t) \right\rangle_0 = \left\langle e^{iHt}|e\rangle_n {}_n\langle g|e^{-iHt}\right\rangle_0 = \left\langle P_{e,g}^{(n)}(t) \right\rangle_0, \tag{4.7a}$$

$$\left\langle \sigma_-^{(n)}(t) \right\rangle_0 = \left\langle e^{iHt}|g\rangle_n {}_n\langle e|e^{-iHt}\right\rangle_0 = \left\langle P_{g,e}^{(n)}(t) \right\rangle_0. \tag{4.7b}$$

However, for the convenience of calculations these quantities should be expressed in terms of a complete set, $|s\rangle$ of the spin system:

$$\sigma_\pm^{(n)}(0) = \sum_{s_1,s_2} c_{\pm,s_1,s_2}^{(n)} |s_1\rangle\langle s_2|. \tag{4.7c}$$

Therefore, for $\left\langle \sigma_\pm^{(n)}(\tau)\right\rangle_0$ we obtain:

$$\left\langle \sigma_\pm^{(n)}(\tau) \right\rangle_0 = \sum_{s_1,s_2} c_{\pm,s_1,s_2}^{(n)} \left\langle P_{s_1,s_2}(\tau) \right\rangle_0. \tag{4.7d}$$

A more non-trivial task is to calculate the two-time correlation functions like the expression in the (4.2). As is seen from (4.2), in order to find a radiation spectrum of our system one needs to calculate the two-time correlation functions of atomic operators. These calculations usually require not only a solution for a density matrix of the whole system but also the calculation of the transition probability distribution. To overcome this, one can use the quantum fluctuation-regression theorem [37, 38], which under some conditions allows for finding two-time correlation functions in terms of one-time correlation functions. However, this theorem still requires to know a complete solution for the density matrix with fixed initial conditions and also usually requres to perform some additional calculations as well [40].

In this regard, we propose here to use a different approach to calculate correlation functions by using the transition operator.

From definition (3.2) we obtain:

$$\left\langle S_k^+(\tau) S_k^-(\tau') \right\rangle = \sum_{n,m}^N g_k^{(m)} g_k^{*(n)} e^{ik(x_n-x_m)} \left\langle \sigma_+^{(n)}(\tau) \sigma_-^{(m)}(\tau') \right\rangle. \tag{4.8}$$

The expression for two-time correlator $\left\langle \sigma_+^{(n)}(\tau)\sigma_-^{(m)}(\tau')\right\rangle$ can be written in two ways depending on the relation between $\tau$ and $\tau'$. If $\tau > \tau'$, then:

$$\left\langle \sigma_+^{(n)}(\tau)\sigma_-^{(m)}(\tau')\right\rangle = Tr_{S,V}\left[\sigma_+^{(n)}(\tau)\sigma_-^{(m)}(\tau')\rho(0)\right] = Tr_{S,V}\left[\rho(\tau')\sigma_+^{(n)}(\tau-\tau')\sigma_-^{(m)}(0)\right], \tag{4.9a}$$

If $\tau < \tau'$, then:

$$\left\langle \sigma_+^{(n)}(\tau)\sigma_-^{(m)}(\tau')\right\rangle = Tr_{S,V}\left[\sigma_+^{(n)}(\tau)\sigma_-^{(m)}(\tau')\rho(0)\right] = Tr_{S,V}\left[\sigma_+^{(m)}(0)\sigma_-^{(n)}(\tau'-\tau)\rho(\tau)\right], \tag{4.9b}$$

These prescriptions come from the requirement for the time argument of a spin operator to be always positive.

The expressions (4.9a), (4.9b) are exact. If we assume the system is *always* in a photon vacuum state:

$$\rho(t) = \rho_S(t) \times \rho_V(0) = \rho_S(t) \times |0\rangle\langle 0|, \tag{4.10}$$

we obtain for (4.9a), (4.9b):

$$\left\langle \sigma_+^{(n)}(\tau)\sigma_-^{(m)}(\tau')\right\rangle = Tr_S\left[\rho_S(\tau')\left\langle\sigma_+^{(n)}(\tau-\tau')\right\rangle_0 \sigma_-^{(m)}(0)\right], \qquad \tau > \tau', \tag{4.11a}$$

$$\left\langle \sigma_+^{(n)}(\tau)\sigma_-^{(m)}(\tau')\right\rangle = Tr_S\left[\sigma_+^{(m)}(0)\left\langle\sigma_-^{(n)}(\tau'-\tau)\right\rangle_0 \rho_S(\tau)\right], \qquad \tau < \tau', \tag{4.11b}$$

For correlation function (4.2) we finally obtain:

$$\langle a_k^\dagger(t)a_k(t)\rangle = \int_0^t d\tau \int_0^\tau d\tau' e^{-i\omega_k(\tau-\tau')}\langle S_k^+(\tau)S_k^-(\tau')\rangle + \int_0^t d\tau \int_\tau^t d\tau' e^{-i\omega_k(\tau-\tau')}\langle S_k^+(\tau)S_k^-(\tau')\rangle$$

$$= \sum_{n,m}^N g_k^{(m)} g_k^{*(n)} e^{ik(x_n-x_m)} \int_0^t d\tau \int_0^\tau d\tau' e^{-i\omega_k(\tau-\tau')} Tr_S\left[\rho_S(\tau')\langle\sigma_+^{(n)}(\tau-\tau')\rangle_0 \sigma_-^{(m)}(0)\right] \quad (4.12)$$

$$+\sum_{n,m}^N g_k^{(m)} g_k^{*(n)} e^{ik(x_n-x_m)} \int_0^t d\tau \int_\tau^t d\tau' e^{-i\omega_k(\tau-\tau')} Tr_S\left[\sigma_+^{(n)}(0)\langle\sigma_-^{(m)}(\tau'-\tau)\rangle_0 \rho_S(\tau)\right]$$

The atomic operators $\langle\sigma_\pm^{(n)}(\tau-\tau')\rangle_0$ in (4.12) should be expressed in terms of transition operators similar to (4.7d).

With the aid of (2.13) we can rewrite the density matrix $\rho_S(t)$ in (4.12) in terms of the initial density matrix $\rho_S(0)$ and transition operators. Finally, for the quantities $Tr_S[\ldots]$ in (4.12) we obtain:

$$Tr_S\left[\rho_S(\tau')\langle\sigma_+^{(n)}(\tau-\tau')\rangle_0 \sigma_-^{(m)}(0)\right] = \sum_{l,q}\langle l|\rho_S(\tau')|q\rangle\langle q|\langle\sigma_+^{(n)}(\tau-\tau')\rangle_0 \sigma_-^{(m)}(0)|l\rangle$$

$$= \sum_{l,q}\sum_{s,p} \langle s|\rho_S(0)|p\rangle\langle p|\langle P_{ql}(\tau')\rangle_0|s\rangle\langle q|\langle\sigma_+^{(n)}(\tau-\tau')\rangle_0 \sigma_-^{(m)}(0)|l\rangle \quad (4.13a)$$

$$Tr_S\left[\sigma_+^{(n)}(0)\langle\sigma_-^{(m)}(\tau'-\tau)\rangle_0 \rho_S(\tau)\right] = \sum_{l,q}\langle l|\sigma_+^{(n)}(0)\langle\sigma_-^{(m)}(\tau'-\tau)\rangle_0|q\rangle\langle q|\rho_S(\tau)|l\rangle$$

$$= \sum_{l,q}\sum_{s,p} \langle l|\sigma_+^{(n)}(0)\langle\sigma_-^{(m)}(\tau'-\tau)\rangle_0|q\rangle\langle s|\rho_S(0)|p\rangle\langle p|\langle P_{lq}(\tau)\rangle_0|s\rangle \quad (4.13b)$$

Therefore, we can calculate the desired correlation functions using only transition operators averaged over the vacuum field.

Note that this approach allows one to calculate not only two-time correlation functions similar to (4.9) but more complex ones as well. In (4.13) we reduce the averaged value of two operators to an average of one operator multiplied by the second operator at zero time. The same principle can be used to reduce, say, a four-time correlation function to a three-time correlation function, which can be also reduced the same way to a two-time correlation function, and so on. Thus, we can find higher-order correlation functions using only transition operators found from (3.11).

## 5. One-qubit system

In this section, mainly for tutorial purpose, we apply the transition operator approach to a single qubit, embedded in a one-dimensional infinite waveguide. This simple model allows one to understand all advantages and possibilities of the transition operator approach. Here we derive the matrix elements for the transition operator for the one-qubit system and apply them to calculate the transition probabilities and the radiation spectrum.

### 5.1 Transition operator for one qubit

For one qubit in a waveguide (N = 1) the basis states of the system will be represented by two states, $|e\rangle$ and $|g\rangle$, corresponding to an excited state and a ground state of a qubit, respectively. If we take the location of the qubit at the center of coordinates ($x_1 = 0$), then the equation (3.11) becomes:

$$\frac{d}{dt}P_{ij} = i\frac{\Omega}{2}\left(e^{iHt}\sigma_Z|i\rangle\langle j|e^{-iHt} - e^{iHt}|i\rangle\langle j|\sigma_Z e^{-iHt}\right)$$

$$+i\sum_k a_k^\dagger(0)e^{i\omega_k t}g_k e^{iHt}\left[\sigma_-,|i\rangle\langle j|\right]e^{-iHt} + i\sum_k g_k^* e^{-i\omega_k t}e^{iHt}\left[\sigma_+,|i\rangle\langle j|\right]e^{-iHt}a_k(0), \quad (5.1)$$

$$+\Gamma e^{iHt}\sigma_+|i\rangle\langle j|\sigma_- e^{-iHt} - \frac{\Gamma}{2}e^{iHt}\sigma_+\sigma_-|i\rangle\langle j|e^{-iHt} - \frac{\Gamma}{2}e^{iHt}|i\rangle\langle j|\sigma_+\sigma_- e^{-iHt}$$

where i, j = e, g, and we introduce a single-qubit decay rate:

$$\Gamma = 2\pi\sum_k |g_k|^2 \delta(\omega_k - \Omega). \quad (5.2a)$$

The frequency shift α

$$\alpha = \sum_k |g_k|^2 P.v.\left(\frac{1}{\omega_k - \Omega}\right). \quad (5.2b)$$

will be included in the renormalized qubit frequency $\Omega$.

Specifying the states i, j and finally returning to the notation of transition operators we end up with the following three equations:

$$\frac{dP_{ee}}{dt} = -\Gamma P_{ee}(t) + i\sum_k a_k^\dagger(0) g_k \, e^{i\omega_k t} P_{ge}(t) - iP_{eg}(t) \sum_k g_k^* \, e^{-i\omega_k t} a_k(0), \qquad (5.3a)$$

$$\frac{dP_{gg}}{dt} = \Gamma P_{ee}(t) - i\sum_k a_k^\dagger(0) g_k \, e^{i\omega_k t} P_{ge}(t) + iP_{eg}(t) \sum_k g_k^* \, e^{-i\omega_k t} a_k(0), \qquad (5.3b)$$

$$\frac{dP_{eg}}{dt} = \left(i\Omega - \frac{\Gamma}{2}\right) P_{eg}(t) + i\sum_k a_k^\dagger(0) g_k \, e^{i\omega_k t} \left(P_{gg}(t) - P_{ee}(t)\right). \qquad (5.3c)$$

It is easy to see that equations (5.3) are linear, i.e. only linear terms containing the transition operators remain on the right hand side. Note also that there are four total transition operators, but the last two are conjugate to each other ($P_{ge}^* = P_{eg}$), so we need to calculate only one of them.

Since we consider a system initially with no photons, equations (5.3) should be averaged over the vacuum state $|0\rangle$:

$$\frac{d}{dt}\langle P_{ee}\rangle_0 = -\Gamma \langle P_{ee}\rangle_0, \qquad (5.4a)$$

$$\frac{d}{dt}\langle P_{gg}\rangle_0 = \Gamma \langle P_{ee}\rangle_0, \qquad (5.4b)$$

$$\frac{d}{dt}\langle P_{eg}\rangle_0 = \left(i\Omega - \frac{\Gamma}{2}\right)\langle P_{eg}\rangle_0. \qquad (5.4c)$$

Taking into account the initial values for the transition operator, $P_{ij}(0) = |i\rangle\langle j|$, the solution of equations (5.4) is given by:

$$\langle P_{ee}(t)\rangle_0 = e^{-\Gamma t}|e\rangle\langle e|, \qquad (5.5a)$$

$$\langle P_{gg}(t)\rangle_0 = 1 - e^{-\Gamma t}|e\rangle\langle e|, \qquad (5.5b)$$

$$\langle P_{eg}(t)\rangle_0 = e^{\left(i\Omega - \frac{\Gamma}{2}\right)t}|e\rangle\langle g|. \qquad (5.5c)$$

We note once again that in contrast to similar equations for the well-known elements of the density matrix, the obtained values (5.5) are the operator functions.

## 5.2 Transition probabilities for one qubit

Now, using the obtained expressions of the transition operators (5.5), we can calculate the transition probabilities for one qubit. Using relation (2.7) we find the transition probabilities for the initially excited qubit:

$$W_{e\to e} = \langle e|\langle P_{ee}\rangle_0|e\rangle = e^{-\Gamma t}, \qquad (5.6a)$$

$$W_{e\to g} = \langle e|\langle P_{gg}\rangle_0|e\rangle = 1 - e^{-\Gamma t}, \qquad (5.6b)$$

If the qubit was initially in the ground state, the probabilities of its transition to an excited and ground states are zero and unity, respectively:

$$W_{g\to e} = \langle g|\langle P_{ee}\rangle_0|g\rangle = 0, \qquad W_{g\to g} = \langle g|\langle P_{gg}\rangle_0|g\rangle = 1. \qquad (5.7)$$

Let's also consider a qubit, which initially is in symmetric superposition:

$$|s\rangle = \frac{1}{\sqrt{2}}|e\rangle + \frac{1}{\sqrt{2}}|g\rangle. \qquad (5.8)$$

Then the probability that after time *t* it remains in the same state is (see equation (2.2)):

$$W_{s\to s} = \langle s|\langle e^{iHt}|s\rangle\langle s|e^{-iHt}\rangle_0|s\rangle = \frac{1}{2}\langle s|\langle P_{ee} + P_{gg} + P_{eg} + P_{ge}\rangle_0|s\rangle$$
$$= \frac{1}{2}\left(1 + e^{-\frac{\Gamma}{2}t} \cos\Omega t\right) \qquad (5.9)$$

Similarly, we can find the probability that a qubit from the superposition state (5.8) will go into an excited state:

$$W_{s \to e} = \frac{1}{2} e^{-\frac{\Gamma}{2}t}, \qquad (5.10)$$

or that it will go to the ground state:

$$W_{s \to g} = 1 - \frac{1}{2} e^{-\frac{\Gamma}{2}t}. \qquad (5.11)$$

The results (5.6-5.11) represent the dynamics of spontaneous emission, in which the initially excited atom exponentially decays into the ground state.

## 5.3 Radiation spectrum for one qubit

The radiation spectrum (4.12) for one qubit takes the following form:

$$\langle a_k^\dagger(t) a_k(t) \rangle = |g_k|^2 \int_0^t d\tau \int_0^\tau d\tau' e^{-i\omega_k(\tau-\tau')} Tr_S \left( \langle \sigma_+(\tau)\sigma_-(\tau') \rangle_0 \rho_S(0) \right)$$
$$+ |g_k|^2 \int_0^t d\tau \int_\tau^t d\tau' e^{-i\omega_k(\tau-\tau')} Tr_S \left( \langle \sigma_+(\tau)\sigma_-(\tau') \rangle_0 \rho_S(0) \right) \qquad (5.12)$$

The calculation of the two-time correlation function $\langle \sigma_+(\tau)\sigma_-(\tau') \rangle$ for both $\tau > \tau'$ and $\tau < \tau'$ was done in Sec.4, and generally takes the form of (4.13). To specify the solution, we need to express atomic lowering and raising operators in terms of transition operators $P_{ij}(t)$. From the definitions $\sigma_- = |g\rangle\langle e|$, $\sigma_+ = |e\rangle\langle g|$ and (2.5), it is easy to show that in the Heisenberg representation they can be written down in the following form:

$$\sigma_+(t) = e^{-iHt} |e\rangle\langle g| e^{iHt} = P_{eg}(t), \qquad \sigma_-(t) = e^{-iHt} |g\rangle\langle e| e^{iHt} = P_{ge}(t). \qquad (5.13)$$

Now using the general formula for two-time correlation function (4.13) and (5.13), for one qubit we obtain:

$$\langle \sigma_+(\tau)\sigma_-(\tau') \rangle = \langle e|\rho_S(0)|e\rangle\langle e|\langle P_{e,e}(\tau')\rangle_0|e\rangle\langle e|\langle P_{e,g}(\tau-\tau')\rangle_0|g\rangle, \qquad \tau > \tau'. \qquad (5.14a)$$

$$\langle \sigma_+(\tau)\sigma_-(\tau') \rangle = \langle e|\rho_S(0)|e\rangle\langle g|\langle P_{g,e}(\tau'-\tau)\rangle_0|e\rangle\langle e|\langle P_{e,e}(\tau)\rangle_0|e\rangle, \qquad \tau < \tau'. \qquad (5.14b)$$

Substituting in (5.14) the expressions of the transition operators found in (5.5) we come to the following expression:

$$\langle \sigma_+(\tau)\sigma_-(\tau') \rangle = e^{\left(i\Omega - \frac{\Gamma}{2}\right)\tau} e^{-\left(i\Omega + \frac{\Gamma}{2}\right)\tau'} \langle e|\rho_S(0)|e\rangle, \qquad (5.15)$$

which is the same for both time intervals. Finally, using (5.15) in (5.12) we obtain a spectrum:

$$\langle a_k^\dagger(t) a_k(t) \rangle = |g_k|^2 \frac{\left(e^{(i\omega_k - i\Omega - \Gamma/2)t} - 1\right)\left(e^{-(i\omega_k - i\Omega + \Gamma/2)t} - 1\right)}{(\omega_k - \Omega)^2 + \Gamma^2/4} \langle e|\rho_S(0)|e\rangle. \qquad (5.16)$$

Note that all calculations are done without specifying the initial state $\rho_S(0)$, which is a unique feature of the transition operator approach. It is clearly seen from (5.16) that the emission spectrum is nonzero for the initial state with an excited qubit $\rho_S(0) = |e\rangle\langle e|$ or for the state with arbitrary superposition $\rho_S(0) = (\alpha|e\rangle + \beta|g\rangle) \times (\alpha^*\langle e| + \beta^*\langle g|)$.

If we consider the system for a very long time, which corresponds to setting the time to infinity $t \to \infty$, we obtain the spectral density:

$$S(\omega) = \frac{v_g}{2L} \frac{\Gamma}{(\omega - \Omega)^2 + \Gamma^2/4} \langle e|\rho(0)|e\rangle, \qquad (5.17)$$

where the spontaneous decay rate $\Gamma$ is given in (3.14c).

In addition, we can also calculate the emission rate:

$$W(t) = \frac{d}{dt} \sum_k \langle a_k^\dagger(t) a_k(t) \rangle, \qquad (5.18)$$

which corresponds to the rate of change of total radiation energy. Using the definition of the spectrum (5.12) we can show that:

$$W(t) = |g_{k_0}|^2 \frac{L}{2\pi v_g} \frac{d}{dt} \int_{-\infty}^{+\infty} d\omega_k \int_0^t d\tau \int_0^t d\tau' e^{-i\omega_k(\tau-\tau')} Tr_S\left(\langle \sigma_+(\tau)\sigma_-(\tau')\rangle_0 \rho_S(0)\right)$$ (5.19)

$$= \frac{\Gamma}{2} Tr_S\left(\langle \sigma_+(\tau)\sigma_-(\tau')\rangle_0 \rho_S(0)\right) = \frac{\Gamma}{2} Tr_S\left(\langle P_{ee}(t)\rangle_0 \rho_S(0)\right) = \frac{\Gamma}{2} e^{-\Gamma t} \langle e|\rho_S(0)|e\rangle$$

Now let's consider the one-time average values of single-photon operators. For one qubit we obtain:

$$\langle a_k(t)\rangle = -ig_k \int_0^t d\tau e^{-i\omega_k(t-\tau)} Tr_S\left(\langle \sigma_-(\tau)\rangle_0 \rho_S(0)\right),$$ (5.20a)

$$\langle a_k^\dagger(t)\rangle = ig_k^* \int_0^t d\tau e^{i\omega_k(t-\tau)} Tr_S\left(\langle \sigma_+(\tau)\rangle_0 \rho_S(0)\right).$$ (5.20b)

With the aid of (5.13) and (5.5c) we finally obtain

$$\langle a_k(t)\rangle = -ig_k e^{-i\omega_k t} \frac{e^{(i\omega_k - i\Omega - \Gamma/2)t} - 1}{i(\omega_k - \Omega) - \Gamma/2} \langle e|\rho_S(0)|g\rangle,$$ (5.21a)

$$\langle a_k^\dagger(t)\rangle = -ig_k^* e^{i\omega_k t} \frac{e^{-(i\omega_k - i\Omega + \Gamma/2)t} - 1}{i(\omega_k - \Omega) + \Gamma/2} \langle g|\rho_S(0)|e\rangle.$$ (5.21b)

Formally, the frequency dependence of both spectra $\langle a_k^\dagger(t)a_k(t)\rangle$, $\langle a_k^\dagger(t)\rangle$, $\langle a_k(t)\rangle$ are similar. However, there is a significant difference: if the qubit is initially excited, $\langle e|\rho(0)|e\rangle = 1$, the quantities $\langle a_k^\dagger(t)\rangle$, $\langle a_k(t)\rangle$ are then exactly zero. However, for the superposition state $\alpha|e\rangle + \beta|g\rangle$ both spectra are nonzero: the quantity $\langle a_k^\dagger(t)a_k(t)\rangle$ is proportional to the population of the excited level, while the quantities $\langle a_k^\dagger(t)\rangle$, $\langle a_k(t)\rangle$ are proportional to the product of the populations of both levels.

## 6. Two-qubit system

For two-qubit system there are four basis states:

$$|1\rangle = |gg\rangle; \quad |2\rangle = |ee\rangle; \quad |3\rangle = |ge\rangle; \quad |4\rangle = |eg\rangle.$$ (6.1)

However, we use here the different basis consisting of states $|1\rangle$, $|2\rangle$ and symmetrical and asymmetrical superposition of states $|3\rangle$ and $|4\rangle$, also known as Bell states:

$$|G\rangle = |gg\rangle; \quad |E\rangle = |ee\rangle$$ (6.2a)

$$|S\rangle = \frac{1}{\sqrt{2}}(|ge\rangle + |eg\rangle); \quad |A\rangle = \frac{1}{\sqrt{2}}(|ge\rangle - |eg\rangle);$$ (6.2b)

The advantage of basis states (6.2) over the basis states (6.1) is that the equations of motion for diagonal matrix elements of transition operator are independent on the off-diagonal ones.

Using the definition of lowering and raising operators for the regular basis (6.1), it is easy to show how they act on the new basis states (6.2):

$$\sigma_+^{(1,2)}|G\rangle = \frac{1}{\sqrt{2}}(|S\rangle \mp |A\rangle), \quad \sigma_-^{(1,2)}|G\rangle = 0,$$

$$\sigma_+^{(1,2)}|E\rangle = 0, \quad \sigma_-^{(1,2)}|E\rangle = \frac{1}{\sqrt{2}}(|S\rangle \pm |A\rangle)$$ (6.3a)

$$\sigma_+^{(1,2)}|S\rangle = \frac{1}{\sqrt{2}}|E\rangle, \quad \sigma_-^{(1,2)}|S\rangle = \frac{1}{\sqrt{2}}|G\rangle$$

$$\sigma_+^{(1,2)}|A\rangle = \pm\frac{1}{\sqrt{2}}|E\rangle, \quad \sigma_-^{(1,2)}|A\rangle = \mp\frac{1}{\sqrt{2}}|G\rangle$$ (6.3b)

The same can be easily done for Pauli spin operators:

$$\sigma_Z^{(1,2)}|G\rangle = -|G\rangle, \quad \sigma_Z^{(1,2)}|E\rangle = |E\rangle,$$
$$\sigma_Z^{(1,2)}|A\rangle = \mp|S\rangle, \quad \sigma_Z^{(1,2)}|S\rangle = \mp|A\rangle. \tag{6.4}$$

## 6.1 Transition operators for two-qubit system

From general equation (3.11), we obtain for N = 2:

$$\begin{aligned}\frac{dP_{ij}}{dt} &= i\frac{\Omega}{2}e^{iHt}\left(\left[\sigma_z^{(1)},|i\rangle\langle j|\right]+\left[\sigma_z^{(2)},|i\rangle\langle j|\right]\right)e^{-iHt}\\&+i\sum_k a_k^\dagger(0)g_k\, e^{i\omega_k t}e^{iHt}\left(e^{ikd/2}\left[\sigma_-^{(1)},|i\rangle\langle j|\right]+e^{-ikd/2}\left[\sigma_-^{(2)},|i\rangle\langle j|\right]\right)e^{-iHt}+\\&+i\sum_k g_k^*\, e^{-i\omega_k t}e^{iHt}\left(e^{-ikd/2}\left[\sigma_+^{(1)},|i\rangle\langle j|\right]+e^{ikd/2}\left[\sigma_+^{(2)},|i\rangle\langle j|\right]\right)e^{-iHt}a_k(0)\\&+\frac{\Gamma}{2}e^{iHt}\left(2\sigma_+^{(1)}|i\rangle\langle j|\sigma_-^{(1)}-\sigma_+^{(1)}\sigma_-^{(1)}|i\rangle\langle j|-|i\rangle\langle j|\sigma_+^{(1)}\sigma_-^{(1)}\right)e^{-iHt}\\&+\frac{\Gamma}{2}e^{iHt}\left(2\sigma_+^{(2)}|i\rangle\langle j|\sigma_-^{(2)}-\sigma_+^{(2)}\sigma_-^{(2)}|i\rangle\langle j|-|i\rangle\langle j|\sigma_+^{(2)}\sigma_-^{(2)}\right)e^{-iHt}\\&+\frac{\Gamma}{2}\cos(k_0 d)\,e^{iHt}\left(2\sigma_+^{(1)}|i\rangle\langle j|\sigma_-^{(2)}-\sigma_+^{(1)}\sigma_-^{(2)}|i\rangle\langle j|-|i\rangle\langle j|\sigma_+^{(1)}\sigma_-^{(2)}\right)e^{-iHt}\\&+\frac{\Gamma}{2}\cos(k_0 d)\,e^{iHt}\left(2\sigma_+^{(2)}|i\rangle\langle j|\sigma_-^{(1)}-\sigma_+^{(2)}\sigma_-^{(1)}|i\rangle\langle j|-|i\rangle\langle j|\sigma_+^{(2)}\sigma_-^{(1)}\right)e^{-iHt}\\&+i\frac{\Gamma}{2}\sin(k_0 d)e^{iHt}\left(\left[\sigma_+^{(1)}\sigma_-^{(2)},|i\rangle\langle j|\right]+\left[\sigma_+^{(2)}\sigma_-^{(1)},|i\rangle\langle j|\right]\right)e^{-iHt}\end{aligned} \tag{6.5}$$

where we used the following notations for the decay rates, $\Gamma_{11} = \Gamma_{22} = \Gamma$, $\Gamma_{12} = \Gamma_{21} = \Gamma\cos(k_0 d)$, and for the frequency shifts $\alpha_{11} = \alpha_{22} = 0$, $\alpha_{12} = \alpha_{21} = \Gamma\sin(k_0 d)/2$ where d is the distance between two qubits.

By averaging the equation (6.5) over the photon vacuum state $|0\rangle$, the terms including photon operators in the second and third lines in (6.5) will be dropped out, and we can obtain equations for the individual transition operators. For the basis (6.2) we have sixteen equations in total, but since the off-diagonal transition operators $P_{i,j}$ is a hermitian conjugate, $P_{j,i} = P_{i,j}^\dagger$, it is sufficient to find the solution only for ten matrix elements of the transition operator. For the diagonal matrix elements of transition operator (which we will refer to as populations by analogy with diagonal elements of density matrix), we find:

$$\frac{d\langle P_{GG}\rangle_0}{dt} = \Gamma(1+\cos k_0 d)\langle P_{SS}\rangle_0 + \Gamma(1-\cos k_0 d)\langle P_{AA}\rangle_0, \tag{6.6a}$$

$$\frac{d\langle P_{EE}\rangle_0}{dt} = -2\Gamma\langle P_{EE}\rangle_0, \tag{6.6b}$$

$$\frac{d\langle P_{SS}\rangle_0}{dt} = \Gamma(1+\cos k_0 d)\langle P_{EE}\rangle_0 - \Gamma(1+\cos k_0 d)\langle P_{SS}\rangle_0, \tag{6.6c}$$

$$\frac{d\langle P_{AA}\rangle_0}{dt} = \Gamma(1-\cos k_0 d)\langle P_{EE}\rangle_0 - \Gamma(1-\cos k_0 d)\langle P_{AA}\rangle_0, \tag{6.6d}$$

The summation of these equations leads to $\frac{d}{dt}\left(\langle P_{EE}\rangle_0 + \langle P_{GG}\rangle_0 + \langle P_{SS}\rangle_0 + \langle P_{AA}\rangle_0\right) = 0$ which is consistent with the evident relation that follows from the completeness of the basis set (6.2).

$$\langle P_{EE}\rangle_0 + \langle P_{GG}\rangle_0 + \langle P_{SS}\rangle_0 + \langle P_{AA}\rangle_0 = \langle e^{iHt}\left(|E\rangle\langle E|+|G\rangle\langle G|+|S\rangle\langle S|+|A\rangle\langle A|\right)e^{-iHt}\rangle_0 = 1$$

For the off-diagonal matrix elements of the transition operator (which we will refer to as coherences) we obtain:

$$\frac{d\langle P_{GE}\rangle_0}{dt} = -(2i\Omega+\Gamma)\langle P_{GE}\rangle_0, \tag{6.7a}$$

$$\frac{d\langle P_{AS}\rangle_0}{dt} = -\Gamma(1+i\sin k_0 d)\langle P_{AS}\rangle_0, \qquad (6.7b)$$

$$\frac{d\langle P_{AE}\rangle_0}{dt} = -i\left(\Omega + \frac{\Gamma}{2}\sin k_0 d\right)\langle P_{AE}\rangle_0 - \frac{\Gamma}{2}(3-\cos k_0 d)\langle P_{AE}\rangle_0, \qquad (6.7c)$$

$$\frac{d\langle P_{SE}\rangle_0}{dt} = -i\left(\Omega - \frac{\Gamma}{2}\sin k_0 d\right)\langle P_{SE}\rangle_0 - \frac{\Gamma}{2}(3+\cos k_0 d)\langle P_{SE}\rangle_0, \qquad (6.7d)$$

$$\frac{d\langle P_{GA}\rangle_0}{dt} = -i\left(\Omega - \frac{\Gamma}{2}\sin k_0 d\right)\langle P_{GA}\rangle_0 - \Gamma(1-\cos k_0 d)\langle P_{AE}\rangle_0 - \frac{\Gamma}{2}(1-\cos k_0 d)\langle P_{GA}\rangle_0, \qquad (6.7e)$$

$$\frac{d\langle P_{GS}\rangle_0}{dt} = -i\left(\Omega + \frac{\Gamma}{2}\sin k_0 d\right)\langle P_{GS}\rangle_0 + \Gamma(1+\cos k_0 d)\langle P_{SE}\rangle_0 - \frac{\Gamma}{2}(1+\cos k_0 d)\langle P_{GS}\rangle_0, \qquad (6.7f)$$

Thus in the basis (6.2), the equations for populations are decoupled from those for the coherences. Moreover, first four equations for the coherences are fully independent and related only to their corresponding matrix elements. These equations can be solved without any problems since the initial conditions, which are based on the definition of transition operator (2.5), are always unique: $P_{ij}(0) = |i\rangle\langle j|$.

By solving two groups of equations we find all matrix elements for the transition operator for a two-qubit system in an open waveguide. For the populations we obtain the following solutions:

$$\langle P_{EE}(t)\rangle_0 = e^{-2\Gamma t}|E\rangle\langle E|, \qquad (6.8a)$$

$$\langle P_{SS}(t)\rangle_0 = |S\rangle\langle S|e^{-\Gamma_+ t} - \frac{1+\cos k_0 d}{1-\cos k_0 d}\left(e^{-2\Gamma t} - e^{-\Gamma_+ t}\right)|E\rangle\langle E|, \qquad (6.8b)$$

$$\langle P_{AA}(t)\rangle_0 = |A\rangle\langle A|e^{-\Gamma_- t} - \frac{1-\cos k_0 d}{1+\cos k_0 d}\left(e^{-2\Gamma t} - e^{-\Gamma_- t}\right)|E\rangle\langle E|, \qquad (6.8c)$$

$$\langle P_{GG}(t)\rangle_0 = |G\rangle\langle G| - \left(e^{-\Gamma_+ t}-1\right)|S\rangle\langle S| - \left(e^{-\Gamma_- t}-1\right)|A\rangle\langle A|$$
$$+ \frac{(1+\cos k_0 d)^2}{1-\cos k_0 d}\left[\frac{(e^{-2\Gamma t}-1)}{2} - \frac{(e^{-\Gamma_+ t}-1)}{1+\cos k_0 d}\right]|E\rangle\langle E| + \frac{(1-\cos k_0 d)^2}{1+\cos k_0 d}\left[\frac{(e^{-2\Gamma t}-1)}{2} - \frac{(e^{-\Gamma_- t}-1)}{1-\cos k_0 d}\right]|E\rangle\langle E|. \qquad (6.8d)$$

For coherences we obtain:

$$\langle P_{GE}(t)\rangle_0 = e^{-(2i\Omega+\Gamma)t}|G\rangle\langle E|, \qquad (6.9a)$$

$$\langle P_{AS}(t)\rangle_0 = e^{-\Gamma(1+i\sin(k_0 d))t}|A\rangle\langle S|, \qquad (6.9b)$$

$$\langle P_{AE}(t)\rangle_0 = e^{-\left(i\Omega_+ + \frac{\Gamma_-}{2}+\Gamma\right)t}|A\rangle\langle E|, \qquad (6.9c)$$

$$\langle P_{SE}(t)\rangle_0 = e^{-\left(i\Omega_- + \frac{\Gamma_+}{2}+\Gamma\right)t}|S\rangle\langle E|, \qquad (6.9d)$$

$$\langle P_{GA}(t)\rangle_0 = e^{-\left(i\Omega_- + \frac{\Gamma_-}{2}\right)t}|G\rangle\langle A| + \frac{1-\cos k_0 d}{1+i\sin k_0 d}\left(e^{-\left(i\Omega_+ + \frac{\Gamma_-}{2}+\Gamma\right)t} - e^{-\left(i\Omega_- + \frac{\Gamma_-}{2}\right)t}\right)|A\rangle\langle E|, \qquad (6.9e)$$

$$\langle P_{GS}(t)\rangle_0 = e^{-\left(i\Omega_+ + \frac{\Gamma_+}{2}\right)t}|G\rangle\langle S| - \frac{1+\cos k_0 d}{1-i\sin k_0 d}\left(e^{-\left(i\Omega_- + \frac{\Gamma_+}{2}+\Gamma\right)t} - e^{-\left(i\Omega_+ + \frac{\Gamma_+}{2}\right)t}\right)|S\rangle\langle E|. \qquad (6.9f)$$

Here for simplification, we introduce the shifted resonant frequencies and modified decay rates:

$$\Omega_+ = \Omega + \frac{\Gamma}{2}\sin k_0 d; \qquad \Omega_- = \Omega - \frac{\Gamma}{2}\sin k_0 d; \qquad (6.10a)$$

$$\Gamma_+ = \Gamma(1+\cos k_0 d); \qquad \Gamma_- = \Gamma(1-\cos k_0 d); \qquad (6.10b)$$

which depends on the effective distance between the qubits.

Unlike the usual solution for the density matrix, expressions (6.8) and (6.9) are the operator functions. Nevertheless, knowing the expressions for the matrix elements of the transition operator, we can easily find the density matrix using relations (2.12) or (2.13).

*6.2 Transition probabilities for two-qubit system*

As noted in Sec.2, the probability of a system to transit from one state to another can be found with the aid of transition operators (see Eq. 2.7). Let us calculate all possible transition probabilities between the basis states (6.2).

*6.2.1 Both qubits are initially excited.*

For both qubits initially in an excited state $|\Psi_0\rangle = |ee\rangle = |E\rangle$. We can then find the probability that at time $t$ the system remains in the initial state:

$$W_{E \to E} = \langle E | P_{EE} | E \rangle = e^{-2\Gamma t} . \tag{6.11a}$$

The probability for both qubits to decay to a ground state is:

$$W_{E \to G} = \langle E | P_{GG} | E \rangle = \frac{(1+\cos k_0 d)^2}{1-\cos k_0 d}\left[\frac{\left(e^{-2\Gamma t}-1\right)}{2} - \frac{\left(e^{-\Gamma(1+\cos k_0 d)t}-1\right)}{1+\cos k_0 d}\right]$$
$$+ \frac{(1-\cos k_0 d)^2}{1+\cos k_0 d}\left[\frac{\left(e^{-2\Gamma t}-1\right)}{2} - \frac{\left(e^{-\Gamma(1-\cos k_0 d)t}-1\right)}{1-\cos k_0 d}\right] . \tag{6.11b}$$

The probability for both qubits to decay to a symmetric Bell state is:

$$W_{E \to S} = \langle E | P_{SS} | E \rangle = -\frac{1+\cos k_0 d}{1-\cos k_0 d}\left(e^{-2\Gamma t} - e^{-\Gamma(1+\cos k_0 d)t}\right), \tag{6.11c}$$

And the probability for both qubits to decay to an asymmetric Bell state is:

$$W_{E \to A} = \langle E | P_{AA} | E \rangle = -\frac{1-\cos k_0 d}{1+\cos k_0 d}\left(e^{-2\Gamma t} - e^{-\Gamma(1-\cos k_0 d)t}\right). \tag{6.11d}$$

Due to the completeness of the basis set (6.2), the sum of all probabilities (6.11a)-(6.11d) is equal to unity.

One can also calculate the probability that only one of the qubits decays to the ground state:

$$W_{E \to ge} = \langle E | e^{iHt} | ge \rangle \langle ge | e^{-iHt} | E \rangle = \frac{1}{2} \langle E | (P_{SS} + P_{AA} + P_{SA} + P_{AS}) | E \rangle =$$
$$= -\frac{1}{2}\left[\frac{1+\cos k_0 d}{1-\cos k_0 d}\left(e^{-2\Gamma t} - e^{-\Gamma_+ t}\right) + \frac{1-\cos k_0 d}{1+\cos k_0 d}\left(e^{-2\Gamma t} - e^{-\Gamma_- t}\right)\right], \tag{6.11e}$$

Moreover, this probability does not depend on the final state $|ge\rangle$ or $|eg\rangle$: the transition to both of these states is equally probable, $W_{E \to eg} = W_{E \to ge}$.

The frequency dependence of different transitions from the state $|E\rangle$ for $k_0 d = \pi/4$ is shown in Fig. 1. The probability for two qubits to remain in excited states (6.11a) exponentially decays with rate $2\Gamma$ and tends to zero for $t \to \infty$, and, consequently, the probability to find both qubits in the ground state (6.11b) tends to 1 (see Fig. 1, red and blue lines). The probabilities of transitions to entangled states $|A\rangle$ and $|S\rangle$, as well as to a one excited qubit state, start from zero, reach some maximum value, and then slowly decays to zero as well (Fig.1, all other lines).

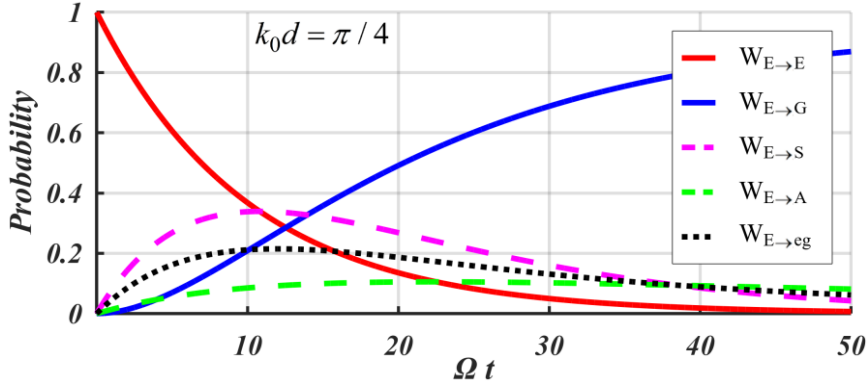

**Figure 1.** Transition probabilities for the initial state in which both qubits are excited. Red line is the probability that qubits will remain excited; blue line is the probability that both qubits will decay to the ground state; purple dashed line, the probability of transition to a symmetrical entangled state; green dashed line, the probability of transition to the asymmetric entangled state; The black dotted line is the probability of transition to the state with one excited qubit; The calculations have been performed for decay rate $\Gamma/\Omega = 0.05$.

It also should be noted, that transitions (6.11) heavily depend on the effective distance between the qubits $k_0 d$ (except for $W_{E \to E}$). For example, for $k_0 d = \pi/2$ there are equal probabilities of transitions to symmetric and asymmetric Bell states, and to the state with one excited qubit: $W_{E \to S} = W_{E \to A} = W_{E \to ge}$.

We should separately consider the case when $k_0 d = n\pi$. For this case, $\cos(k_0 d) = \pm 1$ and both the numerator and denominator in some terms in (6.11b)-(6.11e) tend to zero. The right solution can be obtained if we put $k_0 d = n\pi$ directly in the equations (6.6) and (6.7), or by expanding $\cos(k_0 d)$ near $k_0 d - n\pi$ as a small value. Both approaches give the same result anyway. For $k_0 d = 2n\pi$ we obtain the following transition probabilities:

$$W_{E \to G} = \left(1 - e^{-2\Gamma t}\right) - 2\Gamma t e^{-2\Gamma t}, \tag{6.12a}$$

$$W_{E \to S} = 2\Gamma t e^{-2\Gamma t}, \qquad W_{E \to A} = 0, \tag{6.12b}$$

$$W_{E \to ge} = \frac{1}{2} W_{E \to S} = \Gamma t e^{-2\Gamma t}. \tag{6.12c}$$

As it is clearly seen from (6.12b), for an even number of n = 0, 2, 4…, the transition from state $|E\rangle$ to asymmetric entangled state $|A\rangle$ is forbidden and has zero probability. It is also worth noting that the probability of transition to a symmetric entangled state is exactly two times larger than the probability of transition to a state with only one excited qubit. For an odd number of n = 1, 3… the situation is reversed: the transition to symmetric Bell state is now forbidden, and for asymmetric state we get the relation $W_{E \to A} = 2\Gamma t e^{-2\Gamma t} = 2W_{E \to ge}$.

*6.2.2 Symmetrical entangled state.*

When qubits are initially in a symmetrical Bell state $|\Psi_0\rangle = (|ge\rangle + |eg\rangle)/\sqrt{2} = |S\rangle$, the probability of the system to remain in this state is given by:

$$W_{S \to S} = \langle S | P_{SS} | S \rangle = e^{-\Gamma(1 + \cos k_0 d)t}. \tag{6.14a}$$

The probability to decay to a common ground state is:

$$W_{S \to G} = \langle S | P_{GG} | S \rangle = 1 - e^{-\Gamma(1 + \cos k_0 d)t}. \tag{6.14b}$$

The probability of qubits to became both excited, or switching to an asymmetrical entangled state is zero:

$$W_{S \to E} = W_{S \to A} = 0. \tag{6.14c}$$

Finally, the transition probability from a symmetrical entangled state to a state with only one excited qubit is half the $W_{S \to S}$:

$$W_{S \to ge} = W_{S \to eg} = \frac{1}{2} e^{-\Gamma(1+\cos k_0 d)t}. \tag{6.14d}$$

If we take $k_0 d = \pi$, the cosine will be equal to minus 1, and hence the probability (6.14a) becomes equal to 1. This means that for that particular phase relation the symmetric entangled state is stable and did not decay to a ground state, thus there are no photons that can appear in a waveguide. For that reason, the state $|S\rangle$ for $k_0 d = \pi$ is often called the dark state (we consider this effect in more detail when we derive the spectrum in sec. 6.3.1). Nevertheless, the nonradiative decay of entangled states to single-qubit excited states is possible, and it has equal probabilities no matter what qubit is excited in the final state.

### 6.2.3 Asymmetrical entangled state.

A very similar picture we find for the transition probabilities for an asymmetrical Bell state $|\Psi_0\rangle = (|ge\rangle - |eg\rangle)/\sqrt{2} = |A\rangle$. The probability of the initial state to remain in the same state is:

$$W_{A \to A} = \langle A | P_{AA} | A \rangle = e^{-\Gamma(1-\cos k_0 d)t}. \tag{6.15a}$$

Probability of decay to a ground state:

$$W_{A \to G} = \langle A | P_{GG} | A \rangle = 1 - e^{-\Gamma(1-\cos k_0 d)t}. \tag{6.15b}$$

The probability of qubits to became both excited, or changing asymmetric state to symmetric is zero:

$$W_{A \to E} = W_{A \to S} = 0. \tag{6.15c}$$

And, for transition to a state with one excited qubit:

$$W_{A \to ge} = W_{A \to eg} = \frac{1}{2} e^{-\Gamma(1-\cos k_0 d)t}. \tag{6.15d}$$

As is seen from (6.14c) and (6.15c) the $|A\rangle$ and $|S\rangle$ states are completely decoupled from each other no matter what is the value of $k_0 d$. In addition, for $k_0 d = 2n\pi$, where n is a positive integer, the $|A\rangle$ becomes a dark state as the transitions from $|A\rangle$ to all basis states (6.2) are forbidden. In this case, the state $|A\rangle$ does not interact with the electromagnetic field. If $k_0 d = (2n+1)\pi$ the situation is reversed: the $|S\rangle$ state becomes a dark state.

The symmetric and asymmetric entangled states have different decay rates which depend on the value of $k_0 d$. As is shown in Fig.2 for a given value of $k_0 d$ the decay rates for $|S\rangle$ state are greater than those for $|A\rangle$ state.

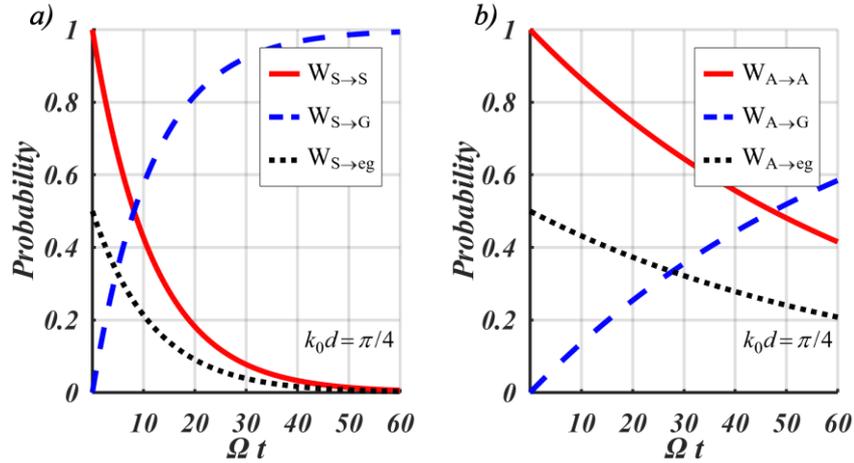

**Figure 2.** Transition probabilities for a two-qubit system with the initial state being symmetric (a) and asymmetric (b) entangled (Bell) state. Red line is the probability of qubits to remain in the initial state; blue dashed line, the probability of decay to the ground state; black dotted line, the probability of transition to the state with only one excited qubit; the other transitions have a probability equals to zero. Both graphs are very similar and differ only by time scale. The decay rate is $\Gamma/\Omega = 0.05$.

*6.2.4 Only one qubit is excited.* Let us now consider a state when only a second qubit is initially excited, $|\Psi_0\rangle = |3\rangle = |ge\rangle = (|S\rangle + |A\rangle)/\sqrt{2}$. In this case, we obtain:

$$W_{ge \to i} = \frac{1}{2}\left(\langle S| + \langle A|\right) P_{ii} \left(|S\rangle + |A\rangle\right). \tag{6.16}$$

The probability of the system making a transition to an entangled state (symmetrical or asymmetrical) is then:

$$W_{ge \to S} = \frac{1}{2} e^{-\Gamma(1+\cos k_0 d)t}, \qquad W_{ge \to A} = \frac{1}{2} e^{-\Gamma(1-\cos k_0 d)t}. \tag{6.17a}$$

The probability of both qubits becoming excited is equal to zero $W_{ge \to E} = 0$, and the probability to decay into the ground state is:

$$W_{ge \to G} = 1 - \frac{1}{2} e^{-\Gamma(1+\cos k_0 d)t} - \frac{1}{2} e^{-\Gamma(1-\cos k_0 d)t}, . \tag{6.17b}$$

Note that relations (6.17a), (6.17b) are also true for the left qubit being excited, i.e. when $|\Psi_0\rangle = |eg\rangle$.

Now let us calculate the probability of excitation to transfer from one qubit to the other one, $|ge\rangle \to |eg\rangle$:

$$W_{ge \to eg} = \frac{1}{4}\left(e^{-\Gamma(1+\cos k_0 d)t} + e^{-\Gamma(1-\cos k_0 d)t} - 2e^{-\Gamma t}\cos\left(\Gamma\sin(k_0 d)t\right)\right). \tag{6.17c}$$

Similarly, the probability that the initially excited second qubit remains in the same state is:

$$W_{ge \to ge} = \frac{1}{4}\left(e^{-\Gamma(1+\cos k_0 d)t} + e^{-\Gamma(1-\cos k_0 d)t} + 2e^{-\Gamma t}\cos\left(\Gamma\sin(k_0 d)t\right)\right). \tag{6.17d}$$

The last terms in (6.17c) and (6.17d) are damped vacuum Rabi oscillations. They transform to regular exponential decay $2e^{-\Gamma t}$ for $k_0 d = n\pi$, and get the maximum frequency for $k_0 d = (n+1/2)\pi$. Again, we get the same relations regardless of what qubit was initially excited.

Transition probabilities for the initial state in which only the second qubit is excited are shown in Fig.3. If we take the arbitrary value of $k_0 d$, say $\pi/4$ (Fig 3a), all probabilities are exponentially decaying to zero at larger times (except for the transition to the ground state). There is also a shallow maximum for the probability to transfer the excitation between qubits. A different picture is observed for $k_0 d = n\pi$ (Fig3b). First, one of the entangled states becomes stable and did not decay at all. For the odd number of n, the stable state is symmetrical, and for the even number of n, it is asymmetrical. Second, for large times the probabilities $W_{ge \to ge}$ and $W_{ge \to eg}$ tend not to a zero, but to a constant value equal to 1/4.

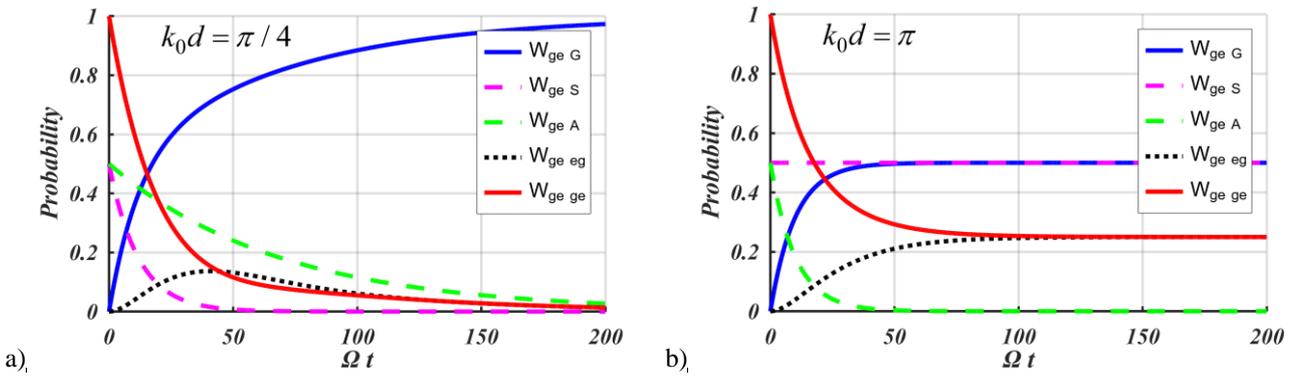

**Figure 3.** Transition probabilities for the initial state in which only the second qubit is excited. The red line is the probability of the initial state to remain in the same state; the blue line, the probability of both qubits to decay to the ground state; purple dashed line, the probability of transition to a symmetrical entangled state; the green dashed line, the probability of transition to an asymmetrical entangled state; the black dotted line, the probability to transfer the excitation to the first qubit. Parameters of the system: $\Gamma/\Omega = 0.05$, a) $k_0 d = \pi/4$, b) $k_0 d = \pi$. For the second case probabilities of one-qubit excitation will tend not to zero, but to 0.25.

For the effective distance $k_0 d = (n+1/2)\pi$ the probabilities for transitions to symmetric and asymmetric Bell states become equal: $W_{ge \to S} = W_{ge \to A} = e^{-\Gamma t}/2$. For the probabilities to transfer the excitation between qubits we obtain from (6.17c) and (6.17d): $W_{ge \to ge} = 0.5 e^{-\Gamma t}(1+\cos(\Gamma t))$ and $W_{ge \to eg} = 0.5 e^{-\Gamma t}(1-\cos(\Gamma t))$. Therefore, in this case, the frequency of the damped Rabi oscillations is a maximum and equals to spontaneous decay rate $\Gamma$. However, the Rabi oscillations in Fig.3a are not explicitly seen as the fast decay rate completely damps them out within one oscillation period.

*6.3 Radiation spectra of two qubits in a waveguide*

Now we switch to the calculation of radiation spectrum for a two-qubit system. As was shown in (4.2), the spectrum can be found using a set of atomic correlation functions. For N = 2 we obtain:

$$\langle a_k^\dagger(t) a_k(t) \rangle = \langle a_k^\dagger(t) a_k(t) \rangle_+ + \langle a_k^\dagger(t) a_k(t) \rangle_- , \quad (6.18a)$$

$$\langle a_k^\dagger(t) a_k(t) \rangle_+ = |g_k|^2 \int_0^t d\tau \int_0^\tau d\tau' e^{-i\omega(\tau-\tau')} \Big[ \langle \sigma_+^{(1)}(\tau) \sigma_-^{(1)}(\tau') \rangle + \langle \sigma_+^{(2)}(\tau) \sigma_-^{(2)}(\tau') \rangle \\ + e^{-ikd} \langle \sigma_+^{(1)}(\tau) \sigma_-^{(2)}(\tau') \rangle + e^{ikd} \langle \sigma_+^{(2)}(\tau) \sigma_-^{(1)}(\tau') \rangle \Big] , \quad (6.18b)$$

$$\langle a_k^\dagger(t) a_k(t) \rangle_- = |g_k|^2 \int_0^t d\tau \int_\tau^t d\tau' e^{-i\omega(\tau-\tau')} \Big[ \langle \sigma_+^{(1)}(\tau) \sigma_-^{(1)}(\tau') \rangle + \langle \sigma_+^{(2)}(\tau) \sigma_-^{(2)}(\tau') \rangle \\ + e^{-ikd} \langle \sigma_+^{(1)}(\tau) \sigma_-^{(2)}(\tau') \rangle + e^{ikd} \langle \sigma_+^{(2)}(\tau) \sigma_-^{(1)}(\tau') \rangle \Big] . \quad (6.18c)$$

Here we subdivide the whole spectrum into two parts, for positive time difference (when $\tau > \tau'$) and negative time difference (when $\tau < \tau'$). It is needed to clarify the future calculations.

To find complete spectra (6.18a) one should calculate the two-time correlation functions using (4.13). In Section 5 for one-qubit system, we used the simple relation (5.13) between $P_{eg}(t)$ and $\sigma_+(t)$. However, for the two-qubit system, the connection between atomic operators and a transition operator is more complex. From the conditions (6.3) we can express the lowering and raising spin operators in terms of basis states (6.2):

$$\sigma_+^{(1,2)} = \frac{1}{\sqrt{2}} \Big[ |S\rangle\langle G| \mp |A\rangle\langle G| + |E\rangle\langle S| \pm |E\rangle\langle A| \Big], \quad (6.19a)$$

$$\sigma_-^{(1,2)} = \frac{1}{\sqrt{2}} \Big[ |G\rangle\langle S| \mp |G\rangle\langle A| + |S\rangle\langle E| \pm |A\rangle\langle E| \Big]. \quad (6.19b)$$

And, by switching to a Heisenberg picture, we find that:

$$\sigma_+^{(1,2)}(t) = \big[ P_{SG}(t) \mp P_{AG}(t) + P_{ES}(t) \pm P_{EA}(t) \big] / \sqrt{2} , \quad (6.20a)$$

$$\sigma_-^{(1,2)}(t) = \big[ P_{GS}(t) \mp P_{GA}(t) + P_{SE}(t) \pm P_{AE}(t) \big] / \sqrt{2} , \quad (6.20b)$$

Thus, one can find the complete spectra (6.18a) by calculating four two-time correlation functions with the already obtained transition operators (6.8) and (6.9).

For the positive time difference $\tau > \tau'$ we will have the following general expression:

$$\left\langle a_k^\dagger(t)a_k(t)\right\rangle_+ = |g_k|^2 \int_0^t d\tau \int_0^t d\tau' e^{-i\omega(\tau-\tau')} \times [$$

$$(1+\cos(kd))\left(\langle E|\langle P_{SG}(\tau-\tau'))\rangle_0|S\rangle\langle E|\langle P_{EE}(\tau'))\rangle_0|E\rangle + \langle E|\langle P_{ES}(\tau-\tau'))\rangle_0|S\rangle\langle E|\langle P_{EE}(\tau'))\rangle_0|E\rangle\right.$$

$$\left.+\langle S|\langle P_{SG}(\tau-\tau'))\rangle_0|G\rangle\langle E|\langle P_{SS}(\tau'))\rangle_0|E\rangle\right)\langle E|\rho_S(0)|E\rangle$$

$$(1-\cos(kd))\left(\langle E|\langle P_{EA}(\tau-\tau'))\rangle_0|A\rangle\langle E|\langle P_{EE}(\tau'))\rangle_0|E\rangle - \langle E|\langle P_{AG}(\tau-\tau'))\rangle_0|A\rangle\langle E|\langle P_{EE}(\tau'))\rangle_0|E\rangle\right.$$

$$\left.+\langle A|\langle P_{AG}(\tau-\tau'))\rangle_0|G\rangle\langle E|\langle P_{AA}(\tau'))\rangle_0|E\rangle\right)\langle E|\rho_S(0)|E\rangle \qquad (6.21a)$$

$$+(1+\cos(kd))\langle S|\langle P_{SG}(\tau-\tau'))\rangle_0|G\rangle\langle S|\langle P_{SS}(\tau'))\rangle_0|S\rangle\langle S|\rho_S(0)|S\rangle$$

$$+(1-\cos(kd))\langle A|\langle P_{AG}(\tau-\tau'))\rangle_0|G\rangle\langle A|\langle P_{AA}(\tau'))\rangle_0|A\rangle\langle A|\rho_S(0)|A\rangle$$

$$+i\sin(kd)\langle A|\langle P_{AG}(\tau-\tau'))\rangle_0|G\rangle\langle A|\langle P_{AS}(\tau'))\rangle_0|S\rangle\langle S|\rho_S(0)|A\rangle$$

$$-i\sin(kd)\langle S|\langle P_{SG}(\tau-\tau'))\rangle_0|G\rangle\langle S|\langle P_{SA}(\tau'))\rangle_0|A\rangle\langle A|\rho_S(0)|S\rangle];$$

And for the negative time difference $\tau < \tau'$ we get a similar expression:

$$\left\langle a_k^\dagger(t)a_k(t)\right\rangle_- = |g_k|^2 \int_0^t d\tau \int_\tau^t d\tau' e^{-i\omega(\tau-\tau')} \times [$$

$$(1+\cos(k_0 d))\left(\langle S|\langle P_{GS}(\tau'-\tau)\rangle_0|E\rangle\langle E|\langle P_{EE}(\tau)\rangle_0|E\rangle + \langle S|\langle P_{SE}(\tau'-\tau)\rangle_0|E\rangle\langle E|\langle P_{EE}(\tau)\rangle_0|E\rangle +\right.$$

$$\left.+\langle G|\langle P_{GS}(\tau'-\tau)\rangle_0|S\rangle\langle E|\langle P_{SS}(\tau)\rangle_0|E\rangle\right)\langle E|\rho_S(0)|E\rangle$$

$$(1-\cos(k_0 d))\left(\langle A|\langle P_{AE}(\tau'-\tau)\rangle_0|E\rangle\langle E|\langle P_{EE}(\tau)\rangle_0|E\rangle - \langle A|\langle P_{GA}(\tau'-\tau)\rangle_0|E\rangle\langle E|\langle P_{EE}(\tau)\rangle_0|E\rangle +\right.$$

$$\left.+\langle G|\langle P_{GA}(\tau'-\tau)\rangle_0|A\rangle\langle E|\langle P_{AA}(\tau)\rangle_0|E\rangle\right)\langle E|\rho_S(0)|E\rangle \qquad (6.21b)$$

$$+(1+\cos(k_0 d))\langle G|\langle P_{GS}(\tau'-\tau)\rangle_0|S\rangle\langle S|\langle P_{SS}(\tau)\rangle_0|S\rangle\langle S|\rho_S(0)|S\rangle$$

$$+(1-\cos(k_0 d))\langle G|\langle P_{GA}(\tau'-\tau)\rangle_0|A\rangle\langle A|\langle P_{AA}(\tau)\rangle_0|A\rangle\langle A|\rho_S(0)|A\rangle$$

$$-i\sin(k_0 d)\langle G|\langle P_{GA}(\tau'-\tau)\rangle_0|A\rangle\langle S|\langle P_{SA}(\tau)\rangle_0|A\rangle\langle A|\rho_S(0)|S\rangle$$

$$+i\sin(k_0 d)\langle G|\langle P_{GS}(\tau'-\tau)\rangle_0|S\rangle\langle A|\langle P_{AS}(\tau)\rangle_0|S\rangle\langle S|\rho_S(0)|A\rangle;$$

Note that these expressions are written in such a way, that they can be used for any initial density matrix of the qubit system, $\rho_S(0)$.

Similar to a one-qubit system, we can also calculate here a total emission rate (5.18). Switching from sum over k to integral over ω in (6.18) we obtain the following expression for W(t):

$$W(t) = \frac{\Gamma}{2}\left[\langle \sigma_+^{(1)}(t)\sigma_-^{(1)}(t)\rangle + \langle \sigma_+^{(2)}(t)\sigma_-^{(2)}(t)\rangle + e^{-ik_0 d}\langle \sigma_+^{(1)}(t)\sigma_-^{(2)}(t)\rangle + e^{ik_0 d}\langle \sigma_+^{(2)}(t)\sigma_-^{(1)}(t)\rangle\right]. \qquad (6.22a)$$

where $\langle \sigma_+^{(i)}(t)\sigma_-^{(j)}(t)\rangle$ can be found as (4.13) but with $\tau = \tau' = t$. Thus, the emission rate can easily be calculated since it is proportional only to single-time correlation functions. With the help of expressions for spin operators (6.20) we can express (6.22a) in terms of transition operators:

$$W(t) = \frac{\Gamma}{2}\left[2\langle E|\langle P_{EE}(t)\rangle_0|E\rangle + (1+\cos(k_0 d))\langle E|\langle P_{SS}(t)\rangle_0|E\rangle + (1-\cos(k_0 d))\langle E|\langle P_{AA}(t)\rangle_0|E\rangle\right]\langle E|\rho_S(0)|E\rangle$$

$$+\frac{\Gamma}{2}(1+\cos(k_0 d))\langle S|\langle P_{SS}(t)\rangle_0|S\rangle\langle S|\rho_S(0)|S\rangle + \frac{\Gamma}{2}(1-\cos(k_0 d))\langle A|\langle P_{AA}(t)\rangle_0|A\rangle\langle A|\rho_S(0)|A\rangle \qquad (6.22b)$$

$$+i\frac{\Gamma}{2}\sin(k_0 d)\left(\langle A|\langle P_{AS}(t)\rangle_0|S\rangle\langle S|\rho_S(0)|A\rangle - \langle S|\langle P_{SA}(t)\rangle_0|A\rangle\langle A|\rho_S(0)|S\rangle\right)$$

This is also a general expression that can be used for any initial density matrix of the qubit system, $\rho_S(0)$.

For a one-time mean value of photon operators in the case of two qubits we obtain

$$\langle a_k(t)\rangle = -ig_k e^{-i\omega t}\int_0^t e^{i\omega\tau}\left(e^{ik_0 d/2}Tr_S\left(\langle \sigma_-^{(1)}(\tau)\rangle_0 \rho_S(0)\right) + e^{-ik_0 d/2}Tr_S\left(\langle \sigma_-^{(2)}(\tau)\rangle_0 \rho_S(0)\right)\right)d\tau, \qquad (6.23a)$$

$$\left\langle a_k^\dagger(t)\right\rangle = \left(\left\langle a_k(t)\right\rangle\right)^\dagger. \qquad (6.23b)$$

Using (6.20b) it is easy to rewrite this in a more suitable form:

$$\left\langle a_k(t)\right\rangle = -i\frac{2}{\sqrt{2}}g_k e^{-i\omega t}\int_0^t d\tau e^{i\omega\tau} \times \Big[\cos(k_0 d/2)\big(\langle S|\langle P_{GS}(\tau)\rangle_0|E\rangle + \langle S|\langle P_{SE}(\tau)\rangle_0|E\rangle\big)\langle E|\rho_S(0)|S\rangle$$
$$+ i\sin(k_0 d/2)\big(\langle A|\langle P_{AE}(\tau)\rangle_0|E\rangle - \langle A|\langle P_{GA}(\tau)\rangle_0|E\rangle\big)\langle E|\rho_S(0)|A\rangle + \qquad (6.24)$$
$$\cos(k_0 d/2)\langle G|\langle P_{GS}(\tau)\rangle_0|S\rangle\langle S|\rho_S(0)|G\rangle - i\sin(k_0 d/2)\langle G|\langle P_{GA}(\tau)\rangle_0|A\rangle\langle A|\rho_S(0)|G\rangle\Big]$$

Note that one-time mean values are proportional only to off-diagonal elements of the initial density matrix.

All expressions presented in (6.21), (6.22b) and (6.24) are written in the general form and can be used for any initial state of the two-qubit system. However, now we will consider specific initial states and calculate the corresponding spectra and other desired parameters. We start with the most common initial states, including Bell's entangled states (symmetric and asymmetric), and states with one and two excited qubits. After that we consider more complex initial states when one or both of the qubits can be prepared in a superposition state. We show that spectra for these types of states can be expressed in terms of the first three obtained common spectra. The main difference, however, is that for superposition states we have non-zero average values of single-photon operators (6.24).

*6.3.1 Initial symmetric and asymmetric Bell states.* We start with initially prepared entangled states in a form of a symmetrical Bell state $|\Psi(0)\rangle = (|e_1 g_2\rangle + |g_1 e_2\rangle)/\sqrt{2} = |S\rangle$ and an asymmetrical Bell state $|\Psi(0)\rangle = (|e_1 g_2\rangle - |g_1 e_2\rangle)/\sqrt{2} = |A\rangle$. The experimental technique for the preparation of these entangled states is widely known in the circuit QED field and can be implemented by the sequence of Hadamard and CNOT gates [5].

For the symmetric initial state, $\rho_S(0) = |S\rangle\langle S|$, we obtain from (6.21):

$$\left\langle a^\dagger(t)a(t)\right\rangle_S = |g_k|^2(1+\cos(kd))\int_0^t d\tau \int_0^\tau d\tau' e^{-i\omega(\tau-\tau')}\langle S|\langle P_{SG}(\tau-\tau')\rangle_0|G\rangle\langle S|\langle P_{SS}(\tau')\rangle_0|S\rangle$$
$$+ |g_k|^2(1+\cos(k_0 d))\int_0^t d\tau\int_\tau^t d\tau' e^{-i\omega(\tau-\tau')}\langle G|\langle P_{GS}(\tau'-\tau)\rangle_0|S\rangle\langle S|\langle P_{SS}(\tau)\rangle_0|S\rangle \qquad (6.25a)$$

Substituting the transition operators found in (6.8) and (6.9) we come to a simple integral expression:

$$\left\langle a^\dagger(t)a(t)\right\rangle_S = |g_k|^2(1+\cos(k_0 d))\int_0^t e^{-(i\delta_+ + \Gamma_+/2)\tau} d\tau \int_0^\tau e^{(i\delta_+ - \Gamma_+/2)\tau'} d\tau'$$
$$+ |g_k|^2(1+\cos(k_0 d))\int_0^t e^{-(i\delta_+ + \Gamma_+/2)\tau} d\tau \int_\tau^t e^{(i\delta_+ - \Gamma_+/2)\tau'} d\tau' \qquad (6.25b)$$

where to simplify the notation, we introduce the detuning parameter

$$\delta_+ = \omega - \Omega_+; \qquad \delta_- = \omega - \Omega_-; \qquad (6.26)$$

Thus, the complete spectrum for symmetrical Bell state is given by:

$$\left\langle a_k^\dagger(t)a_k(t)\right\rangle_S = \frac{\upsilon_g \Gamma_+}{2L}\frac{\left(e^{(i\delta_+ - \Gamma_+/2)t}-1\right)\left(e^{-(i\delta_+ + \Gamma_+/2)t}-1\right)}{\delta_+^2 + \Gamma_+^2/4}. \qquad (6.27a)$$

If now we let time tend to infinity, $t \to \infty$, we get a radiation spectrum that is dependent only on the frequency:

$$S_S(\omega) = \frac{\upsilon_g}{2L}\frac{\Gamma_+}{\left(\delta_+^2 + \Gamma_+^2/4\right)}. \qquad (6.27b)$$

And the last thing we need to find is the total emission rate. From (6.22b) we obtain:

$$W_S(t) = \frac{\Gamma_+}{2}e^{-\Gamma_+ t}. \qquad (6.27c)$$

The calculation process for asymmetric Bell state $\rho_S(0) = |A\rangle\langle A|$ is very similar. For this initial state, we obtain from (6.21):

$$\langle a^{\dagger}(t)a(t)\rangle_A = |g_k|^2 (1-\cos(k_0 d))\int_0^t d\tau \int_0^\tau d\tau' e^{-i\omega(\tau-\tau')} \langle A|\langle P_{AG}(\tau-\tau')\rangle_0 |G\rangle \langle A|\langle P_{AA}(\tau')\rangle_0 |A\rangle$$

$$+|g_k|^2 (1-\cos(k_0 d))\int_0^t d\tau \int_\tau^t d\tau' e^{-i\omega(\tau-\tau')} \langle G|\langle P_{GA}(\tau'-\tau)\rangle_0 |A\rangle \langle A|\langle P_{AA}(\tau)\rangle_0 |A\rangle \qquad (6.28)$$

Thus, for asymmetric superposition, we obtain similar to (6.27) the following expression for the spectrum, spectral density, and emission rate:

$$\langle a_k^{\dagger}(t)a_k(t)\rangle_A = \frac{\upsilon_g \Gamma_-}{2L} \frac{\left(e^{(i\delta_- - \Gamma_-/2)t}-1\right)\left(e^{-(i\delta_- + \Gamma_-/2)t}-1\right)}{\delta_-^2 + \Gamma_-^2/4}. \qquad (6.29a)$$

$$S_A(\omega) = \frac{\upsilon_g}{2L}\frac{\Gamma_-}{\delta_-^2 + \Gamma_-^2/4}. \qquad (6.29b)$$

$$W_A(t) = \frac{\Gamma_-}{2}e^{-\Gamma_- t}. \qquad (6.29c)$$

The radiation spectrum for entangled Bell states is shown in Fig.4 for different values of $k_0 d$. We also should note, that the one-time average values of single-photon operators (6.24) are equal to zero for both entangled states since they are proportional only to off-diagonal elements of initial density matrix.

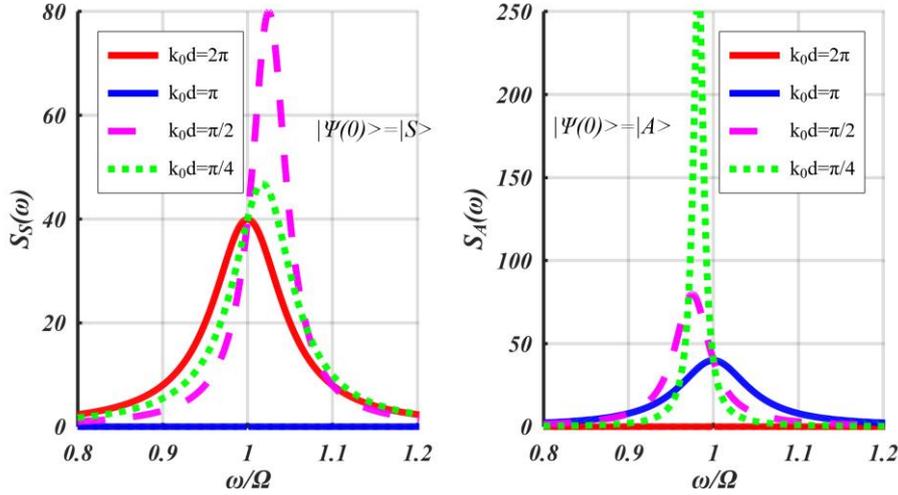

**Figure 4. a)** Radiation spectrum for initial symmetric Bell state for different phase parameter $k_0 d$ (expression (6.27b). When the effective distance between qubits is equal to even numbers of $\pi$ (red line) there is no displacement in frequency, and the spectrum is identical to the one-qubit case but with doubled decay rate $2\Gamma$. On the contrary, when $k_0 d$ is the odd number of $\pi$ (blue line) the spectrum is zero. For the $k_0 d$ proportional to $\pi/2$ (purple dashed line) there is a maximum frequency displacement by the value $(\Omega + \Gamma/2)$ to the right side from the resonant frequency. Intermediate values of $k_0 d$ like $\pi/4$ (green dotted line) will lie in between the first two cases. **b)** Radiation spectrum for the initial asymmetric Bell state, the expression (6.29b). This case is very similar to the symmetric one, but now for $k_0 d = 2\pi$ the spectrum is zero, and for $k_0 d = \pi$ it coincides with a one-qubit spectrum with doubled line width. Also, the displacement of frequency now happens to the left side of the central frequency $(\Omega - \Gamma/2)$. For the $k_0 d = \pi/4$ the line width becomes smaller, as $k_0 d \to 2\pi$, $\Gamma \to 0$. The dimensionless decay rate $\Gamma/\Omega = 0.05$.

The radiation spectrum of a two-qubit system heavily depends on the parameter $k_0 d$, which in some way represents the effective distance between qubits (Fig. 4). For example, let's take $k_0 d = 2\pi$ for the symmetric entangled state (6.27b) which will correspond to the distance $d = \lambda_0$, where $\lambda_0 = 2\pi\upsilon_g/\Omega$ is a wavelength corresponding to the qubit frequency. Then $\cos(k_0 d) = 1$ and a collective decay rate $\Gamma_+$ becomes equal to doubled decay rate of a single qubit: $\Gamma_+ = 2\Gamma$. Thus, the spectrum $S_S(\omega)$ turns out to be identical to the one for a one-qubit system (5.17), but with a doubled line width. This is, in fact, is the manifestation of the Dicke superradiance, when the spectral line width is proportional to the number of atoms in the system, $N = 2$ in our case. On the other hand, if we take the same value $k_0 d = 2\pi$ and look at the spectrum of asymmetric entangled state (6.29b), then we will see that decay rate $\Gamma_- = 0$, and, consequently, the entire spectrum $S_A(\omega)$ equals to zero. Therefore, for an effective distance d = λ₀, the symmetric state $|S\rangle$ is often called bright, since the radiation rate for it is twice as high as compared to a single atom, and the asymmetric state $|A\rangle$ is called dark since it does not radiate at all [21]. It also should be noted that if k₀d is equal to an odd number of π, which corresponds to half the wavelength d = λ₀/2, the picture is reversed: $|S\rangle$ becomes dark, and $|A\rangle$ becomes bright.

For the values of k₀d = (n+1/2)π which corresponds to d = λ₀/4 both symmetric and asymmetric spectra have the same linewidth $\Gamma_+ = \Gamma_- = \Gamma$, but the peaks will be displaced in opposite directions because of the frequency shift $\delta_\pm = \omega - (\Omega \pm \Gamma/2)$ (purple dashed lines on figure 4). This indicates left and a right qubit in a two-qubit system. Going further, when k₀d is tending to 2π for an asymmetric spectrum (or to π for a symmetric), the line width will tend toward zero, which results in a much higher and narrow peak (see green dotted line in fig. 4b).

*6.3.2 Initial state with one excited qubit.* Now we consider the initial state when only the first qubit is excited $|\Psi(0)\rangle = |e_1 g_2\rangle$. The corresponding initial value of the density matrix then:

$$\rho_S(0) = |eg\rangle\langle eg| = \frac{1}{2}|S-A\rangle\langle S-A| = \frac{1}{2}(|S\rangle\langle S| + |A\rangle\langle A| - |S\rangle\langle A| - |A\rangle\langle S|). \tag{6.30}$$

In this case, the general expression (6.21) becomes:

$$\langle a_k^\dagger(t) a_k(t)\rangle_{eg} = \frac{|g_k|^2}{2} \int_0^t d\tau \int_0^\tau d\tau' e^{-i\omega(\tau-\tau')} \times \Big[(1+\cos(k_0 d))\langle S|\langle P_{SG}(\tau-\tau')\rangle_0|G\rangle\langle S|\langle P_{SS}(\tau')\rangle_0|S\rangle$$
$$+ (1-\cos(k_0 d))\langle A|\langle P_{AG}(\tau-\tau')\rangle_0|G\rangle\langle A|\langle P_{AA}(\tau')\rangle_0|A\rangle$$
$$+ i\sin(k_0 d)\langle A|\langle P_{AG}(\tau-\tau')\rangle_0|G\rangle\langle A|\langle P_{AS}(\tau')\rangle_0|S\rangle - i\sin(k_0 d)\langle S|\langle P_{SG}(\tau-\tau')\rangle_0|G\rangle\langle S|\langle P_{SA}(\tau')\rangle_0|A\rangle\Big]$$
$$+ \frac{|g_k|^2}{2} \int_0^t d\tau \int_\tau^t d\tau' e^{-i\omega(\tau-\tau')} \times \Big[(1+\cos(k_0 d))\langle G|\langle P_{GS}(\tau'-\tau)\rangle_0|S\rangle\langle S|\langle P_{SS}(\tau)\rangle_0|S\rangle$$
$$+ (1-\cos(k_0 d))\langle G|\langle P_{GA}(\tau'-\tau)\rangle_0|A\rangle\langle A|\langle P_{AA}(\tau)\rangle_0|A\rangle$$
$$+ i\sin(k_0 d)\langle G|\langle P_{GA}(\tau'-\tau)\rangle_0|A\rangle\langle S|\langle P_{SA}(\tau)\rangle_0|A\rangle - i\sin(k_0 d)\langle G|\langle P_{GS}(\tau'-\tau)\rangle_0|S\rangle\langle A|\langle P_{AS}(\tau)\rangle_0|S\rangle\Big] \tag{6.31}$$

The terms which are proportional to $\langle S|\rho_S(0)|S\rangle$ and $\langle A|\rho_S(0)|A\rangle$ in (6.31) coincide with the expressions (6.25a) and (6.28), except for the factor 1/2. Therefore, we need to calculate only half of (6.31) with the off-diagonal elements of the transition operator. The final expression takes the following form:

$$\langle a_k^\dagger(t) a_k(t)\rangle_{eg} = \frac{1}{2}\langle a_k^\dagger(t) a_k(t)\rangle_S + \frac{1}{2}\langle a_k^\dagger(t) a_k(t)\rangle_A +$$
$$+ i\frac{\upsilon_g \Gamma}{2L} \frac{\sin(k_0 d)}{2} \left[\frac{\left(e^{-(i\delta_- + \Gamma_-/2)t} - 1\right)\left(e^{(i\delta_+ - \Gamma_+/2)t} - 1\right)}{(i\delta_- + \Gamma_-/2)(i\delta_+ - \Gamma_+/2)} - \frac{\left(e^{-(i\delta_+ + \Gamma_+/2)t} - 1\right)\left(e^{(i\delta_- - \Gamma_-/2)t} - 1\right)}{(i\delta_+ + \Gamma_+/2)(i\delta_- - \Gamma_-/2)}\right], \tag{6.32a}$$

where the first two terms are given in (6.27a) and (6.29a). In the limit $t \to \infty$, we find frequency-dependent spectrum density:

$$S_{eg}(\omega) = \frac{1}{2}S_S(\omega) + \frac{1}{2}S_A(\omega) - \frac{\upsilon_g \Gamma}{2L} \frac{\sin(k_0 d)}{2} \frac{\delta_- \Gamma_+ - \delta_+ \Gamma_-}{(\delta_+^2 + \Gamma_+^2/4)(\delta_-^2 + \Gamma_-^2/4)}, \tag{6.32b}$$

where $S_S(\omega)$ and $S_A(\omega)$ are given in (6.27b) and (6.29b). Finally, for the total emission rate we obtain:

$$W_{eg}(t) = \frac{\Gamma_+}{4}e^{-\Gamma_+ t} + \frac{\Gamma_-}{4}e^{-\Gamma_- t} - \frac{\Gamma \sin(k_0 d)}{2}e^{-\Gamma t} \sin(\Gamma \sin(k_0 d) t), \tag{6.32c}$$

where the first two terms, again, corresponds to the emission rate of entangled states $W_S(t)/2$ and $W_A(t)/2$, respectively.

If we switch excited qubit in the initial state, i.e. the second one would be excited, $|\Psi(0)\rangle = |g_1 e_2\rangle$, then the density matrix reads:

$$\rho_S(0) = |g_1 e_2\rangle\langle g_1 e_2| = \frac{1}{2}|S+A\rangle\langle S+A| = \frac{1}{2}(|S\rangle\langle S| + |A\rangle\langle A| + |S\rangle\langle A| + |A\rangle\langle S|). \tag{6.33}$$

This results in the sigh change of the interference term in (6.32a), and we get:

$$\langle a_k^\dagger(t)a_k(t)\rangle_{ge} = \frac{1}{2}\langle a_k^\dagger(t)a_k(t)\rangle_S + \frac{1}{2}\langle a_k^\dagger(t)a_k(t)\rangle_A -$$
$$-i\frac{v_g\Gamma}{2L}\frac{\sin(k_0 d)}{2}\left[\frac{\left(e^{-(i\delta_-+\Gamma_-/2)t}-1\right)\left(e^{(i\delta_+-\Gamma_+/2)t}-1\right)}{(i\delta_-+\Gamma_-/2)(i\delta_+-\Gamma_+/2)} - \frac{\left(e^{-(i\delta_++\Gamma_+/2)t}-1\right)\left(e^{(i\delta_--\Gamma_-/2)t}-1\right)}{(i\delta_++\Gamma_+/2)(i\delta_--\Gamma_-/2)}\right],\quad (6.34a)$$

$$S_{ge}(\omega) = \frac{1}{2}S_S(\omega) + \frac{1}{2}S_A(\omega) + \frac{v_g\Gamma}{2L}\frac{\sin(k_0 d)}{2}\frac{\delta_-\Gamma_+ - \delta_+\Gamma_-}{\left(\delta_+^2+\Gamma_+^2/4\right)\left(\delta_-^2+\Gamma_-^2/4\right)},\quad (6.34b)$$

$$W_{ge}(t) = \frac{\Gamma_+}{4}e^{-\Gamma_+ t} + \frac{\Gamma_-}{4}e^{-\Gamma_- t} + \frac{\Gamma\sin(k_0 d)}{2}e^{-\Gamma t}\sin\left(\Gamma\sin(k_0 d)t\right).\quad (6.34c)$$

The average values of single-photon operators (6.24) are still being zero.

A closer analysis of expression (6.32b) is presented in Fig.5a. For initially one excited qubit there isn't much difference between phases $k_0 d = 2\pi$ and $k_0 d = \pi$ like in the previous case of entangled states. For both effective distances $d = \lambda_0$ and $d = \lambda_0/2$ we get a single-peaked Lorenz line with doubled line width $2\Gamma$. But for $k_0 d = \pi/2$, which corresponds to $d = \lambda_0/4$, the third term in (6.32b) takes maximum value due to $\sin(k_0 d) = 1$. This results in two separate peaks displaced by the value of $\Gamma/2$ from the central frequency $\Omega$, with the line reaching zero at the point of zero detuning. In a general sense, each peak corresponds to an excited qubit, which means the interaction between the qubits is at the maximum level. Thus, the peaks are the same high, which means that the probability of the first qubit radiating a photon and the probability that the first qubit will pass excitation to the second one are identical. By taking intermediate values of $k_0 d$ we can break this symmetry of interaction between the qubits, which results in a much higher peak for one of them and a small one for the other.

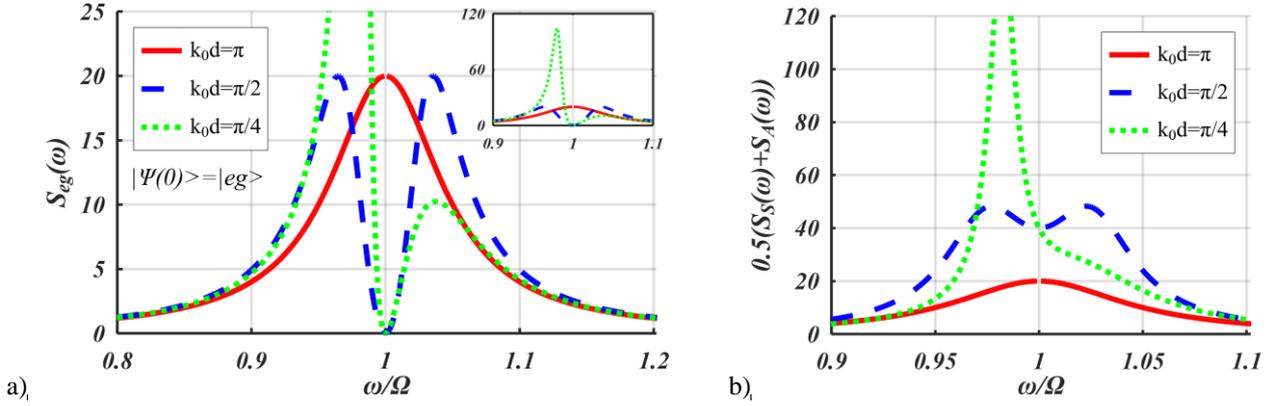

**Figure 5. a)** Spectrum for initial state with one excited qubit (6.34b) for the different effective distances between the qubits. When parameter $k_0 d = n\pi$ (for this initial state there is no difference between $\pi$ and $2\pi$) the spectrum has a single peak on the resonant frequency, similar to a one-qubit case, but with doubled decay rate $2\Gamma$ (red line). When $k_0 d = (n+1/2)\pi$ the interaction takes its maximum and there are two separate peaks, representing the first and second qubits (blue dashed line). When $k_0 d = \pi/4$ (green dotted line), the symmetry between the left and the right peak is breaking, resulting in a much larger left peak (or right if we take $k_0 d = 3\pi/4$). As the value of effective qubit distance will approach an integer number of $\pi$, the frequency displacement moves towards the center frequency. **b)** The same spectrum (6.34b), but without the last interference term, meaning that $S_{eg}(\omega) = 0.5\left(S_S(\omega) + S_A(\omega)\right)$. The spectrum for $k_0 d = n\pi$ (red line) is identical to the previous one. For other values of $k_0 d$ (blue dashed and green dotted lines), the peak separation is not so significant, without the dip to a zero at the central frequency. The decay rate is the same as in Figure 4.

It is also interesting to analyze the spectrum (6.32b) without an interference term when we only have the sum of entangled spectra, $S_{eg}(\omega) = 0.5\left(S_S(\omega) + S_A(\omega)\right)$. It is presented in Fig. 5b. For $k_0 d = n\pi$ nothing changes, because the interference term is proportional to $\sin(k_0 d)$ and it equals zero anyway. However, note that without interference term the spectrum is no longer has the dip to a zero at the central point $\delta = 0$. In other words, the interference term in (6.32b) provides more extensive peak separation between left and right qubits.

*6.3.3 Initial state with two excited qubits.* The last major calculation that we need to perform is the spectrum for the initial state with both qubits being excited, $|\Psi(0)\rangle = |e_1 e_2\rangle = |E\rangle$. The corresponding density matrix is $\rho_S(0) = |E\rangle\langle E|$, and we get from (6.21):

$$\langle a_k^\dagger(t) a_k(t)\rangle_E = \langle a_k^\dagger(t) a_k(t)\rangle_+ + \langle a_k^\dagger(t) a_k(t)\rangle_-, \tag{6.35}$$

where:

$$\langle a_k^\dagger(t) a_k(t)\rangle_+ = |g_k|^2 \int_0^t d\tau \int_0^\tau d\tau' e^{-i\omega(\tau-\tau')} \times \Big[$$
$$(1+\cos(kd))\Big(\langle E|\langle P_{SG}(\tau-\tau')\rangle_0|S\rangle\langle E|\langle P_{EE}(\tau')\rangle_0|E\rangle + \langle E|\langle P_{ES}(\tau-\tau')\rangle_0|S\rangle\langle E|\langle P_{EE}(\tau')\rangle_0|E\rangle$$
$$+ \langle S|\langle P_{SG}(\tau-\tau')\rangle_0|G\rangle\langle E|\langle P_{SS}(\tau')\rangle_0|E\rangle\Big) +$$
$$(1-\cos(kd))\Big(\langle E|\langle P_{EA}(\tau-\tau')\rangle_0|A\rangle\langle E|\langle P_{EE}(\tau')\rangle_0|E\rangle - \langle E|\langle P_{AG}(\tau-\tau')\rangle_0|A\rangle\langle E|\langle P_{EE}(\tau')\rangle_0|E\rangle$$
$$+ \langle A|\langle P_{AG}(\tau-\tau')\rangle_0|G\rangle\langle E|\langle P_{AA}(\tau')\rangle_0|E\rangle\Big)\Big] \tag{6.36a}$$

$$\langle a_k^\dagger(t) a_k(t)\rangle_- = |g_k|^2 \int_0^t d\tau \int_\tau^t d\tau' e^{-i\omega(\tau-\tau')} \times \Big[$$
$$(1+\cos(k_0 d))\Big(\langle S|\langle P_{GS}(\tau'-\tau)\rangle_0|E\rangle\langle E|\langle P_{EE}(\tau)\rangle_0|E\rangle + \langle S|\langle P_{SE}(\tau'-\tau)\rangle_0|E\rangle\langle E|\langle P_{EE}(\tau)\rangle_0|E\rangle +$$
$$+ \langle G|\langle P_{GS}(\tau'-\tau)\rangle_0|S\rangle\langle E|\langle P_{SS}(\tau)\rangle_0|E\rangle\Big) +$$
$$(1-\cos(k_0 d))\Big(\langle A|\langle P_{AE}(\tau'-\tau)\rangle_0|E\rangle\langle E|\langle P_{EE}(\tau)\rangle_0|E\rangle - \langle A|\langle P_{GA}(\tau'-\tau)\rangle_0|E\rangle\langle E|\langle P_{EE}(\tau)\rangle_0|E\rangle +$$
$$+ \langle G|\langle P_{GA}(\tau'-\tau)\rangle_0|A\rangle\langle E|\langle P_{AA}(\tau)\rangle_0|E\rangle\Big)\Big] \tag{6.36b}$$

Substituting the transition operators, we will get the following expression for the spectrum:

$$\langle a_k^\dagger(t) a_k(t)\rangle_E = \frac{\Gamma_+}{\Gamma_-}\langle a_k^\dagger(t) a_k(t)\rangle_S + \frac{\Gamma_-}{\Gamma_+}\langle a_k^\dagger(t) a_k(t)\rangle_A$$
$$+ \frac{v_g}{2L}\frac{\Gamma_+^2 + \Gamma_-^2}{2\Gamma}\left(e^{-2\Gamma t}-1\right)\Bigg[\frac{e^{-ik_0 d}}{\Gamma_+(1+i\sin(k_0 d))(i\delta_- - \Gamma_-/2)} - \frac{e^{-ik_0 d}}{\Gamma_-(1-i\sin(k_0 d))(i\delta_+ - \Gamma_+/2)}$$
$$+ \frac{e^{ik_0 d}}{\Gamma_-(1+i\sin(k_0 d))(i\delta_+ - \Gamma_-/2 - \Gamma)} - \frac{e^{ik_0 d}}{\Gamma_+(1-i\sin(k_0 d))(i\delta_- - \Gamma_+/2 - \Gamma)}\Bigg]$$
$$+ \frac{v_g}{2L\Gamma_+}\frac{e^{ik_0 d}}{(1-i\sin(k_0 d))}\frac{\left(e^{-(i\delta_- + \Gamma_-/2)t}-1\right)\left(\Gamma_-^2 + \Gamma_+^2 e^{(i\delta_- - \Gamma_+/2 - \Gamma)t}\right)}{(i\delta_- + \Gamma_-/2)(i\delta_- - \Gamma_+/2 - \Gamma)}$$
$$- \frac{v_g}{2L\Gamma_-}\frac{e^{ik_0 d}}{(1+i\sin(k_0 d))}\frac{\left(e^{-(i\delta_+ + \Gamma_+/2)t}-1\right)\left(\Gamma_+^2 + \Gamma_-^2 e^{(i\delta_+ - \Gamma_-/2 - \Gamma)t}\right)}{(i\delta_+ + \Gamma_+/2)(i\delta_+ - \Gamma_-/2 - \Gamma)}$$
$$+ \frac{v_g}{2L\Gamma_-}\frac{e^{-ik_0 d}}{(1-i\sin(k_0 d))}\frac{\left(e^{-(i\delta_+ + \Gamma_-/2 + \Gamma)t}-1\right)\left(\Gamma_-^2 + \Gamma_+^2 e^{(i\delta_+ - \Gamma_+/2)t}\right)}{(i\delta_+ - \Gamma_+/2)(i\delta_+ + \Gamma_-/2 + \Gamma)}$$
$$- \frac{v_g}{2L\Gamma_+}\frac{e^{-ik_0 d}}{(1+i\sin(k_0 d))}\frac{\left(e^{-(i\delta_- + \Gamma_+/2 + \Gamma)t}-1\right)\left(\Gamma_+^2 + \Gamma_-^2 e^{(i\delta_- - \Gamma_-/2)t}\right)}{(i\delta_- - \Gamma_-/2)(i\delta_- + \Gamma_+/2 + \Gamma)} \tag{6.37a}$$

And again, by taking time to infinity, we get the spectral density function:

$$S_E(\omega) = \frac{\Gamma_+}{\Gamma_-} S_S(\omega) + \frac{\Gamma_-}{\Gamma_+} S_A(\omega) + \frac{v_g}{2L} \frac{\Gamma_+^2 + \Gamma_-^2}{2\Gamma(1+\sin^2(k_0 d))} \left[ \frac{e^{ik_0 d}(1+i\sin(k_0 d))}{\Gamma_+(i\delta_- - \Gamma_+/2 - \Gamma)} - \frac{e^{-ik_0 d}(1-i\sin(k_0 d))}{\Gamma_+(i\delta_- - \Gamma_-/2)} \right.$$
$$\left. + \frac{e^{-ik_0 d}(1+i\sin(k_0 d))}{\Gamma_-(i\delta_+ - \Gamma_+/2)} - \frac{e^{ik_0 d}(1-i\sin(k_0 d))}{\Gamma_-(i\delta_+ - \Gamma_-/2 - \Gamma)} \right]$$
$$+ \frac{v_g}{2L} \frac{1}{(1+\sin^2(k_0 d))} \left[ \frac{\Gamma_+^2}{\Gamma_+} \frac{e^{-ik_0 d}(1-i\sin(k_0 d))}{(i\delta_- - \Gamma_-/2)(i\delta_- + \Gamma_+/2 + \Gamma)} - \frac{\Gamma_-^2}{\Gamma_+} \frac{e^{ik_0 d}(1+i\sin(k_0 d))}{(i\delta_- + \Gamma_-/2)(i\delta_- - \Gamma_+/2 - \Gamma)} \right]$$
$$+ \frac{v_g}{2L} \frac{1}{(1+\sin^2(k_0 d))} \left[ \frac{\Gamma_+^2}{\Gamma_-} \frac{e^{ik_0 d}(1-i\sin(k_0 d))}{(i\delta_+ + \Gamma_+/2)(i\delta_+ - \Gamma_-/2 - \Gamma)} - \frac{\Gamma_-^2}{\Gamma_-} \frac{e^{-ik_0 d}(1+i\sin(k_0 d))}{(i\delta_+ - \Gamma_+/2)(i\delta_+ + \Gamma_-/2 + \Gamma)} \right]$$
(6.37b)

Much simpler expression we get for emission rate:

$$W_E(t) = \frac{1}{2} \frac{\Gamma_+^2}{\Gamma_-} e^{-\Gamma_+ t} + \frac{1}{2} \frac{\Gamma_-^2}{\Gamma_+} e^{-\Gamma_- t} - \frac{4\Gamma \cos^2(k_0 d)}{1-\cos^2(k_0 d)} e^{-2\Gamma t},$$
(6.37c)

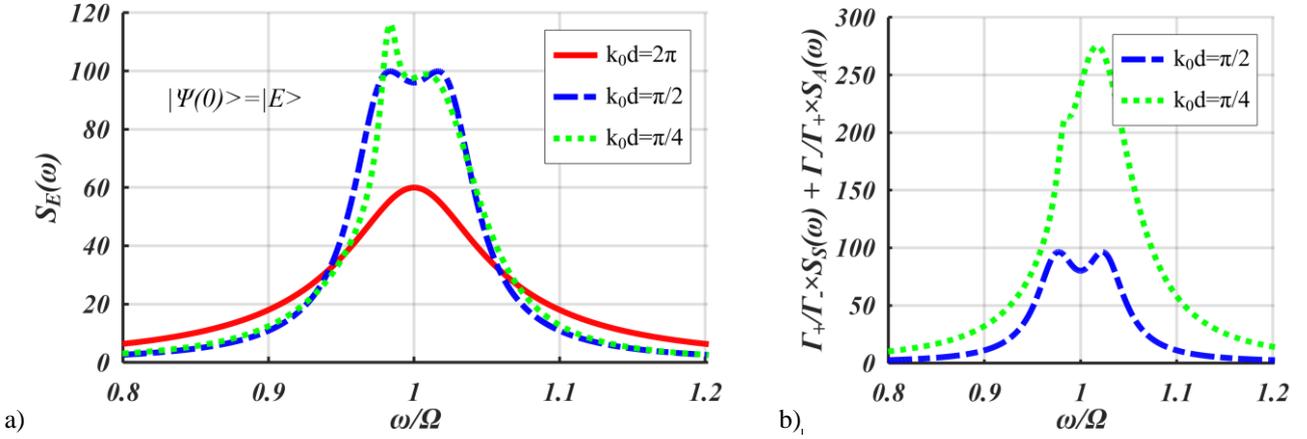

**Figure 6. a)** Spectrum for initial state with two excited qubits (6.37b) for the different effective distances between the qubits. Again, for $k_0 d = \pi$ we have one peak with doubled line width (red line). For $k_0 d = \pi/2$ these peak separates into two peaks, but contrary to figure 5, don't reach zero at the central frequency (blue line). For intermediate values like $k_0 d = \pi/4$ the double-peaked picture becomes asymmetrical. **b)** Incomplete spectrum as the combination of entangled spectra (first two terms of (6.37b)). The main difference is seen for $k_0 d = \pi/4$ (green dotted line). The decay rate is the same as in Figure 4.

One could notice, that expressions (6.37) can get infinite values for $k_0 d = n\pi$. As in the sec. 6.2.1, we can obtain the right solution if we put $k_0 d = n\pi$ directly in the equations (6.6) and (6.7), or by expanding $\cos(k_0 d)$ near $k_0 d = n\pi$ as a small value. For $k_0 d = 2\pi$ we find:

$$\langle a_k^\dagger(t) a_k(t) \rangle_E \bigg|_{k_0 d = 2\pi} = 4 \frac{v_g \Gamma}{2L} \frac{\left(e^{-(i\delta+\Gamma)t}-1\right)\left(e^{(i\delta-\Gamma)t}-1\right)}{\delta^2 + \Gamma^2} + 2 \frac{v_g \Gamma}{2L} \frac{\Gamma^2}{i\delta(i\delta-\Gamma)^2(i\delta-2\Gamma)}$$
$$+ 2 \frac{v_g \Gamma}{2L} \frac{e^{-2\Gamma t}}{(i\delta-\Gamma)^2} - 2 \frac{v_g \Gamma}{2L} \frac{e^{-(i\delta+2\Gamma)t}-1}{i\delta(i\delta+2\Gamma)} - 2 \frac{v_g \Gamma}{2L} \frac{e^{(i\delta-2\Gamma)t}}{i\delta(i\delta-2\Gamma)},$$
(6.38a)
$$- 2 \frac{v_g \Gamma}{2L} \frac{2\Gamma}{(i\delta-\Gamma)^2} \frac{e^{-(i\delta+\Gamma)t}-1}{(i\delta+\Gamma)} + 2 \frac{v_g \Gamma}{2L} \frac{2\Gamma}{(i\delta-\Gamma)} \frac{e^{(i\delta-\Gamma)t}-e^{-2\Gamma t}}{(i\delta+\Gamma)^2} + 2 \frac{v_g \Gamma}{2L} \frac{e^{-2\Gamma t}}{\delta^2 + \Gamma^2} 2\Gamma t$$

$$S_E(\omega)\bigg|_{k_0 d = 2\pi} = \frac{v_g}{2L} \frac{4\Gamma}{\delta^2 + \Gamma^2} + \frac{v_g}{2L} \frac{2\Gamma(\delta^2 - 2\Gamma^2)}{(\delta^2 + \Gamma^2)(\delta^2 + 4\Gamma^2)},$$
(6.38b)

$$W_E(t)|_{k_0d=2\pi} = (1+2\Gamma t)\Gamma e^{-2\Gamma t},  \quad (6.38c)$$

And again, the average values of photon operators (6.24) are zero.

Analysis of the two-excited qubit spectra is done in Fig. 6a. For $k_0d = 2\pi$ we obtain a familiar single-peaked Lorentzian line, though the analytical function (6.38b) is more complex. For $k_0d = \pi/2$ we have two peaks with a small separation at the top, which is similar to the simple sum of entangled spectra presented in Fig.5b but is twice higher. For $k_0d = \pi/4$ we again have symmetry displaced to one of the peaks but without significant growth in values. As in the previous example, we also plot the reduced two-qubit spectrum with only the first two terms in (6.37b), Fig. 6b. For $k_0d = \pi/2$ we have the same line as in Fig.5b because $\frac{\Gamma_+}{\Gamma_-}S_S(\omega) + \frac{\Gamma_-}{\Gamma_+}S_A(\omega) = S_S(\omega) + S_A(\omega)$. For $k_0d = \pi/4$ the spectrum changes drastically, both in values and in peaks proportions – now the right one is dominating instead of the left one.

*6.3.4 Initial states with qubit superposition.* Now we will closely look at the states when qubits are initialized in superposition states. We start with only one excitation when the first qubit is in superposition, and the other one is in a ground state:

$$|\Psi(0)\rangle = |s_1\rangle \otimes |g_2\rangle = \frac{1}{\sqrt{2}}(|e_1\rangle + |g_1\rangle) \otimes |g_2\rangle = \frac{1}{\sqrt{2}}(|e_1g_2\rangle + |g_1g_2\rangle) = \frac{1}{2}|S\rangle - \frac{1}{2}|A\rangle + \frac{1}{\sqrt{2}}|G\rangle. \quad (6.39a)$$

The corresponding density matrix then will be:

$$\rho_S(0) = \frac{1}{4}(|S\rangle\langle S| + |A\rangle\langle A| - |S\rangle\langle A| - |A\rangle\langle S|)$$
$$+ \frac{1}{2\sqrt{2}}(|S\rangle\langle G| - |A\rangle\langle G| + |G\rangle\langle S| - |G\rangle\langle A|) + \frac{1}{2}|G\rangle\langle G| \quad (6.39b)$$

The first line in (6.39b) is matched with a density matrix of the state $|eg\rangle$ (6.30), so we get the same spectrum and other related parameters like for the one excited qubit (6.32), but reduced by the factor of two:

$$\langle a_k^\dagger(t)a_k(t)\rangle_{s_1g_2} = \frac{1}{2}\langle a_k^\dagger(t)a_k(t)\rangle_{eg}; \quad S_{s_1g_2}(\omega) = \frac{1}{2}S_{eg}(\omega); \quad W_{s_1g_2}(t) = \frac{1}{2}W_{eg}(t); \quad (6.40)$$

The second line in (6.39b) whioch describes the transitions to the ground state does not contribute to $\langle a_k^\dagger(t)a_k(t)\rangle$ (5.21).

Therefore the state with the first qubit prepared in a superposition state and the second one in a ground state shows the identical behavior as in Fig.5, but on a smaller scale. However, because there are some off-diagonal elements in (6.39b), for this state there will be non-zero average values of photon operators (6.24):

$$\langle a_k(t)\rangle_{s_1g_2} = -i\frac{g_k}{2}e^{-i\omega t}\left(\cos\left(\frac{k_0d}{2}\right)\frac{e^{(i\delta_+ - \Gamma_+/2)t} - 1}{i\delta_+ - \Gamma_+/2} - i\sin\left(\frac{k_0d}{2}\right)\frac{e^{(i\delta_- - \Gamma_-/2)t} - 1}{i\delta_- - \Gamma_-/2}\right). \quad (6.41)$$

The other average of the creation operator can be found as a complex conjugate of (6.41). The non-zero values of these parameters are unique to initial states that contain superposition.

The next state we will analyze is very similar to the previous one, but now the second qubit is excited:

$$|\Psi(0)\rangle = |s_1\rangle \otimes |e_2\rangle = \frac{1}{\sqrt{2}}(|e_1\rangle + |g_1\rangle) \otimes |e_2\rangle = \frac{1}{\sqrt{2}}|E\rangle + \frac{1}{2}|S\rangle + \frac{1}{2}|A\rangle, \quad (6.42a)$$

and the corresponding density matrix is:

$$\rho_S(0) = \frac{1}{2}|E\rangle\langle E| + \frac{1}{4}(|S\rangle\langle S| + |A\rangle\langle A| + |S\rangle\langle A| + |A\rangle\langle S|)$$
$$+ \frac{1}{2\sqrt{2}}(|E\rangle\langle S| + |E\rangle\langle A| + |S\rangle\langle E| + |A\rangle\langle E|) \quad (6.42b)$$

As in the previous example, the first line in (6.42b) corresponds to some initial states already described above. Therefore, we can construct the spectrum and emission rate using (6.34) and (6.37):

$$\langle a_k^\dagger(t)a_k(t)\rangle_{s_1e_2} = \frac{1}{2}\langle a_k^\dagger a_k\rangle_E + \frac{1}{2}\langle a_k^\dagger a_k\rangle_{ge}; \quad S_{s_1e_2}(\omega) = \frac{1}{2}S_E(\omega) + \frac{1}{2}S_{ge}(\omega);$$
$$W_{s_1e_2}(t) = \frac{1}{2}W_E(t) + \frac{1}{2}W_{ge}(t); \quad (6.43)$$

Thus, the spectrum of state (6.42a) is a combination of the spectrum of two-excited qubits and a one-excited qubit, and it is presented in Fig.7a. The second line in (6.42b) again contributes only to the average values of single-photon operators, which can be found from (6.24):

$$\langle a_k(t) \rangle_{s_1 e_2} = -i \frac{g_k}{2} e^{-i\omega t} \frac{\cos(k_0 d/2)}{1-i\sin(k_0 d)} \left( (1+\cos(k_0 d)) \frac{e^{(i\delta_+ - \Gamma_+/2)t}-1}{i\delta_+ - \Gamma_+/2} - e^{ik_0 d} \frac{e^{(i\delta_- - \Gamma_+/2-\Gamma)t}-1}{i\delta_- - \Gamma_+/2 - \Gamma} \right)$$
$$+ \frac{g_k}{2} e^{-i\omega t} \frac{\sin(k_0 d/2)}{1+i\sin(k_0 d)} \left( (1-\cos(k_0 d)) \frac{e^{(i\delta_- - \Gamma_-/2)t}-1}{i\delta_- - \Gamma_-/2} + e^{ik_0 d} \frac{e^{(i\delta_+ - \Gamma_-/2-\Gamma)t}-1}{i\delta_+ - \Gamma_-/2 - \Gamma} \right)$$
(6.44)

The final state we will consider is when both qubits are prepared in a superposition state:

$$|\Psi(0)\rangle = |s_1\rangle \otimes |s_2\rangle = \frac{1}{\sqrt{2}}(|e_1\rangle + |g_1\rangle) \otimes \frac{1}{\sqrt{2}}(|e_2\rangle + |g_2\rangle) = \frac{1}{2}|E\rangle + \frac{1}{\sqrt{2}}|S\rangle + \frac{1}{2}|G\rangle,$$
(6.45a)

with the density matrix:

$$\rho_S(0) = \frac{1}{4}|E\rangle\langle E| + \frac{1}{2}|S\rangle\langle S| + \frac{1}{4}|E\rangle\langle G| + \frac{1}{4}|G\rangle\langle E| + \frac{1}{4}|G\rangle\langle G|$$
$$+ \frac{1}{2\sqrt{2}}(|E\rangle\langle S| + |S\rangle\langle E| + |S\rangle\langle G| + |G\rangle\langle S|)$$
(6.45b)

This state can be presented as a combination of a two-excited qubit state (6.37) and a symmetrical Bell state (6.27):

$$\langle a_k^\dagger(t) a_k(t) \rangle_{s_1 s_2} = \frac{1}{4}\langle a_k^\dagger a_k \rangle_E + \frac{1}{2}\langle a_k^\dagger a_k \rangle_S; \quad S_{s_1 s_2}(\omega) = \frac{1}{4}S_E(\omega) + \frac{1}{2}S_S(\omega);$$
$$W_{s_1 s_2}(t) = \frac{1}{4}W_E(t) + \frac{1}{2}W_S(t);$$
(6.46)

More close analysis of the spectrum for this state is presented in Fig.7b. For the average values of single-photon operators we get:

$$\langle a_k(t) \rangle_{s_1 s_2} = -i \frac{g_k}{2} e^{-i\omega t} \frac{\cos(k_0 d/2)}{1-i\sin(k_0 d)} \left( (1+\cos(k_0 d)) \frac{e^{(i\delta_+ - \Gamma_+/2)t}-1}{i\delta_+ - \Gamma_+/2} - e^{ik_0 d} \frac{e^{(i\delta_- - \Gamma_+/2-\Gamma)t}-1}{i\delta_- - \Gamma_+/2 - \Gamma} \right)$$
$$-i \frac{g_k}{2} e^{-i\omega t} \cos\left(\frac{k_0 d}{2}\right) \frac{e^{(i\delta_+ - \Gamma_+/2)t}-1}{i\delta_+ - \Gamma_+/2}$$
(6.47)

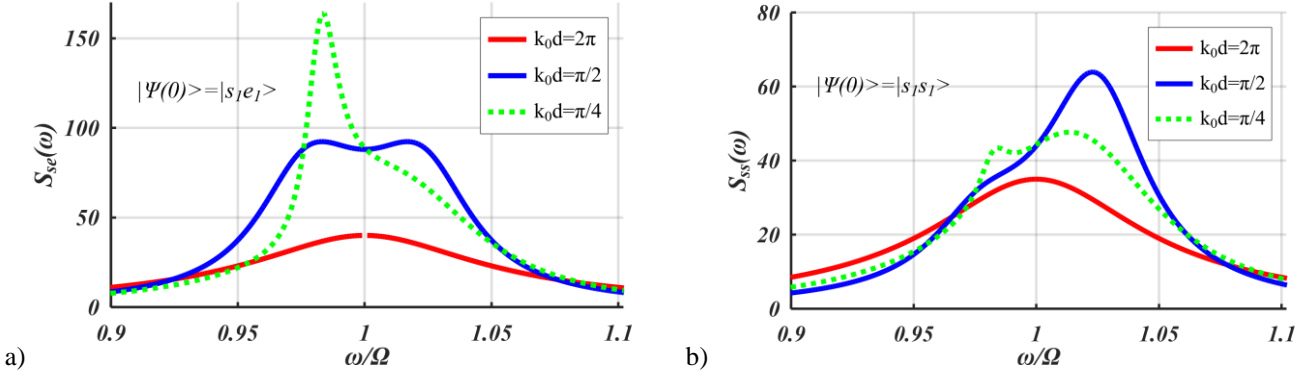

**Figure 7. a)** Spectrum for the initial state with the first qubit being in a superposition state and the second being in the excited state (6.43) for the different $k_0 d$. **b)** Spectrum for initial state with both qubits being in a superposition state (6.46) for the different $k_0 d$. The decay rate is $\Gamma/\Omega = 0.05$.

As can be seen from Fig. 7a, the radiation spectra for the initial state with the first qubit being in a superposition state and the other being in an excited state are very similar to the spectrum of two excited qubits shown in Fig. 6a. On the other hand, when both qubits are prepared in a superposition state, the spectrum changes significantly, and only for $k_0 d = n\pi$ similarity is retained. Note that the line width of both spectrum (6.43) and (6.46) for $k_0 d = 2\pi$ is identical, moreover, it matches with the line width of the spectrum of two excited qubits (6.38b) (red line in Fig. 6a).

# 7. Conclusion

In this paper we show that the transition operator introduced by Lehmberg [33] can be very useful for the theoretical calculations of qubit systems within the circuit QED framework. We successfully apply the transition operator approach for calculating the spontaneous emission spectra for one-qubit and two-qubit systems in a one-dimensional open waveguide. Within the Hilberft space of the spin system the equations of motion for the mattrix elements of the transition operator are linear which allows for the calculation of the two-time correlation functions without recourse to quantum regression theorem. The transition operator allows one to find the transition probabilities within the system and radiation spectra, and overcome some drawbacks of the density matrix approach. As the matrix elements of the transition operator are themselves operators in the Hilbert space of the spin system, in order to get the solution there is no need for the initial state of the spin system to be specified from the very beginning. Theoretically, the transition operator approach allows one to calculate the radiation spectra beyond the one-excitation limit, as we show for some transition probabilities and spectra for the two-qubit system in Sec. 6.

In this paper we generally consider the initial photon state as photon vacuum, focusing mainly on the spontaneous decay of already excited qubits. However, the photon vacuum is not mandatory in the transition operator framework as the general equation (3.11) contains the photon terms that can play a major role for initial states with photon fields.

## Acknowledgments

The work is supported by the Ministry of Science and Higher Education of Russian Federation under the project FSUN-2020-0004 and by the Foundation for the Advancement of Theoretical Physics and Mathematics "BASIS".

## References

[1] Haroche S and Raimond J-M 2006 *Exploring the quantum: atoms, cavities and photons* (Oxford University Press, Oxford)
[2] You J Q and Nori F 2005 Superconducting circuits and quantum information *Phys. Today* **58** (11) 42
[3] Blais A, Gambetta J, Wallraff A, Schuster D I, Girvin S M, Devoret M H and Schoelkopf R J 2007 Quantum information processing with circuit quantum electrodynamics *Phys. Rev. A* **75** 032329
[4] Devoret M H and Schoelkopf R J 2013 Superconducting circuits for quantum information: an outlook *Science* **339** 1169
[5] Wendin G 2017 Quantum information processing with superconducting circuits: a review *Rep. Prog. Phys.* **80** 106001
[6] Neumeier L, Leib M and Hartmann M J 2013 Single-photon transistor in circuit quantum electrodynamics *Phys. Rev. Lett.* **111** 063601
[7] Manzoni M T, Reiter F, Taylor J M and Sørensen A S 2014 Single-photon transistor based on superconducting systems *Phys. Rev. B* **89** 180502(R)
[8] Kyriienko O and Sørensen A S 2016 Continuous-wave single-photon transistor based on superconducting circuit *Phys. Rev. Lett.* **117** 140503
[9] Zueco D, Mazo J J, Solano E, Garcia-Ripoll J J 2012 Microwave photonics with Josephson junction arrays: negative refraction index and entanglement through disorder *Phys. Rev. B* **86** 024503
[10] Macha P, Oelsner G, Reiner J-M, Marthaler M, Andre S, Schon G, Hubner U, Meyer H-G, Il'ichev E and Ustinov A 2014 Implementation of quantum metamaterial using superconducting qubits *Nat. Commun.* **5** 5146
[11] Gu X, Kockum A F, Miranowicz A, Liu Yu-xi and Nori F 2017 Microwave photonics with superconducting quantum circuits *Phys. Rep.* **718-719** 1
[12] Blais A, Huang R-S, Wallraff A, Girvin S M, Schoelkopf R J 2004 Cavity quantum electrodynamics for superconducting electrical circuits: An architecture for quantum computation *Phys. Rev. A* **69** 062320
[13] Wallraff A, Schuster D I, Blais A, Frunzio L, Huang R-S, Majer J, Kumar S, Girvin S M and Schoelkopf R J 2004 Strong coupling of a single photon to a superconducting qubit using circuit quantum electrodynamics *Nature (London)* **431** 162
[14] Clarke J and Wilhelm F K 2008 Superconducting quantum bits *Nature* **453** 1031
[15] Astafiev O, Zagoskin A M, Abdumalikov A A, Pashkin Yu A, Yamamoto T, Inomata K, Nakamura Y, and Tsai J S 2010 Resonance fluorescence of a single artificial atom *Science* **327** 840
[16] You J Q and Nori F 2011 Atomic physics and quantum optics using superconducting circuits *Nature* **474** 589
[17] Forn-Diaz P, Lamata L, Rico E, Kono J and Solano E 2019 Ultrastrong coupling regimes of light-matter interaction *Rev. Mod. Phys.* **91** 025005
[18] Shevchenko S N 2019 *Mesoscopic physics meets quantum engineering* (World Scientific, Singapore)
[19] Ficek Z and Sanders B C 1990 Quantum beats in two-atom resonance fluorescence *Phys. Rev. A* **41** 359
[20] Ordonez G and Kim S 2004 Complex collective states in a one-dimensional two-atom system *Phys. Rev. A* **70** 032702
[21] Lalumiere K, Sanders B C, van Loo A F, Fedorov A, Wallraff A and Blais A 2013 Input-output theory for waveguide QED with an ensemble of inhomogeneous atoms *Phys. Rev. A* **88** 043806
[22] van Loo A F, Fedorov A, Lalumiere K, Sanders B C, Blais A and Wallraff A 2013 Photon-mediated interactions between distant artificial atoms *Science* **342** 1494

[23] Delanty M, Rebic S and Twamley J 2011 Superradiance and phase multistability in circuit quantum electrodynamics *New J. Phys.* **13** 053032

[24] Mlynek J A, Abdumalikov A A, Eichler C and Wallraff A 2014 Observation of Dicke superradiance for two artificial atoms in a cavity with high decay rate *Nat. Commun.* **5** 5186

[25] Lambert N, Matsuzaki Y, Kakuyanagi K, Ishida N, Saito S and Nori F 2016 Superradiance with an ensemble of superconducting flux qubits *Phys. Rev. B* **94** 224510

[26] Albrecht A, Henriet L, Asenjo-Garcia A, Dieterle P B, Painter O and Chang D E 2019 Subradiant states of quantum bits coupled to a one-dimensional waveguide *New J. Phys.* **21** 025003

[27] Zhang Y-X and Molmer K 2019 Theory of subradiant states of a one-dimensional two-level atom chain *Phys. Rev. Lett.* **122** 203605

[28] Koshino K, Kono S and Nakamura Y 2020 Protection of a qubit via subradiance: a Josephson quantum filter *Phys. Rev. Appl.* **13** 014051

[29] Zheng H and Baranger H U 2013 Persistent quantum beats and long-distance entanglement from waveguide-mediated interactions *Phys. Rev. Lett.* **110** 113601

[30] Kannan B, Campbell D L, Vasconcelos F, Winik R, Kim D K, Kjaergaard M, Krantz P, Melville A, Niedzielski B M, Yoder J L, Orlando T P, Gustavsson S, Oliver W D 2020 Generating spatially entangled itinerant photons with waveguide quantum electrodynamics *Sci. Adv.* **6** eabb8780

[31] Xia K, Macovei M and Evers J 2011 Stationary entanglement in strongly coupled qubits *Phys. Rev. B* **84** 184510

[32] Shahmoon E and Kurizki G 2013 Nonradiative interaction and entanglement between distant atoms *Phys. Rev. A* **87** 033831

[33] Lehmberg R H 1969 Transition operators in radiative damping theory *Phys. Rev.* **181** 32

[34] Lehmberg R H 1970 Radiation from an N-Atom system. I. General formalism *Phys. Rev. A* **2** 883

[35] Lehmberg R H 1970 Radiation from an N-atom system. II. Spontaneous emission from a pair of atoms *Phys. Rev. A* **2** 889

[36] Makarov A A, Letokhov V S 2003 Spontaneous decay in a system of two spatially separated atoms (one-dimensional case) *J. Exp. Theor. Phys.* **97** 688–701

[37] Lax M 1963 Formal theory of quantum fluctuations from a driven state *Phys. Rev.* **129** 2342

[38] Scully M O and Zubairy M S 1997 *Quantum Optics* (Cambridge University Press, Cambridge) pp 300-302

[39] Greenberg Y S, Shtygashev A A and Moiseev A G 2021 Spontaneous decay of artificial atoms in a three-qubit system *Eur. Phys. J. B* **94** 221

[40] Gardiner C W and Zoller P 2000 *Quantum noise: A handbook of Markovian and non-Markovian quantum stochastic methods with applications to quantum optics* (Springer, Berlin) pp 369